\documentclass[12pt]{article}

\usepackage{amsfonts}
\usepackage{amsmath}
\usepackage{epsfig}
\usepackage{latexsym}
\usepackage{graphicx}
\usepackage{amssymb}
\usepackage{color}
\numberwithin{equation}{section}
\usepackage{url}
\allowdisplaybreaks

\def\spa#1{\phantom{\fbox{\rule[-#1cm]{0cm}{0cm}}}}

\def\be{\begin{equation}}
\def\ee{\end{equation}}
\def\bea{\begin{eqnarray}}
\def\eea{\end{eqnarray}}
\def\bequ{\begin{equation}}
\def\eequ{\end{equation}}

\def\del{\partial}

\renewcommand{\thefootnote}{\fnsymbol{footnote}}

\newcommand{\eq} {equation}
\newcommand{\eqa} {eqnarray}
\newcommand{\NN} {\mbox {$\nonumber$}}
\newcommand{\eps} {\epsilon}

\newcommand{\bw} {\bar{w}}
\newcommand{\bz} {\bar{z}}

\newcommand{\tsigma} {\tilde{\sigma}}
\newcommand{\tzeta} {\tilde{\zeta}}
\newcommand{\dalpha} {\dot{\alpha}}
\newcommand{\dbeta} {\dot{\beta}}

\newcommand{\calA}{\mathcal{A}}
\newcommand{\caltA}{\tilde{\mathcal{A}}}
\newcommand{\calF}{\mathcal{F}}
\newcommand{\calL}{\mathcal{L}}
\newcommand{\lam}{\lambda}
\newcommand{\tlam}{\tilde{\lambda}}

\newcommand{\tbeta}{\tilde{\beta}}
\newcommand{\tgam}{\tilde{\gamma}}
\newcommand{\tphi}{\tilde{\phi}}
\newcommand{\tpsi}{\tilde{\psi}}
\newcommand{\tF}{\tilde{F}}
\newcommand{\tildea}{\tilde{a}}
\newcommand{\tildeb}{\tilde{b}}
\newcommand{\tildec}{\tilde{c}}
\newcommand{\tildef}{\tilde{f}}

\newcommand{\br}{\mathbf{r}}

\begin{document}

\hfuzz=100pt
\title{Supersymmetric index on $T^2 \times S^2$ and elliptic genus}
\date{}
\author{Masazumi Honda$^{a}$\footnote{masazumihondaAThri.res.in}$\ $  
and Yutaka Yoshida$^{b}$\footnote{yyyyosidaATgmail.com}
  \spa{0.5} \\
\\
$^a${\small{\it Harish-Chandra Research Institute,}}
\\ {\small{\it Chhatnag Road, Jhusi, Allahabad 211019, India}} \\
$^b${\small {\it School of Physics, Korea Institute for Advanced Study (KIAS),}}
\\ {\small {\it 85 Hoegiro Dongdaemun-gu, Seoul, 130-722, Korea}} 
}
\date{\small{April 2015}}

\maketitle
\thispagestyle{empty}
\centerline{}

\begin{abstract}
We study partition function of four-dimensional $\mathcal{N}=1$ supersymmetric field theory
on $T^2 \times S^2$.
By applying supersymmetry localization,
we show that
the $T^2 \times S^2$ partition function is given by elliptic genus of certain two-dimensional $\mathcal{N}=(0,2)$ theory.
As an application,
we discuss a relation between 4d Seiberg duality duality and 2d $(0,2)$ triality,
proposed by Gadde, Gukov and Putrov.
In other examples,
we identify 4d theories giving elliptic genera
of K3, M-strings and E-strings.
In the example of K3,
we find that
there are two 4d theories giving the elliptic genus of K3.
This would imply new four-dimensional duality.
\end{abstract}

\renewcommand{\thefootnote}{\arabic{footnote}}
\setcounter{footnote}{0}

\newpage
\setcounter{page}{1}
\tableofcontents

\section{Introduction}
Two-dimensional conformal field theories (CFTs) have infinite dimensional symmetry
while higher dimensional conformal symmetry is finite dimensional.
This is one of reasons why higher dimensional CFTs are less under control compared to the 2d CFTs.
It would be nice
if some classes of higher dimensional CFTs are related to 2d CFTs in some ways.
In this paper,
we illustrate that
some observables in 4d $\mathcal{N}=1$ supersymmetric gauge theory 
at infrared fixed point are described by 2d CFTs. 
Specifically we study partition function of 4d $\mathcal{N}=1$ supersymmetric gauge theory 
with R-symmetry on $T^2 \times S^2$.
When we impose appropriate boundary conditions,
this partition function is interpreted as 
the following supersymmetric index \cite{Closset:2013sxa}
\begin{\eq}
Z_{T^2 \times S^2} = {\rm Tr}\Bigl[ (-1)^F q^P x^{J_3} \prod_a t_a^{F_a} \Bigr] ,
\label{eq:4dindex}
\end{\eq}
where $F$ is Fermion number, $P$ is momentum along spatial $S^1$ of $T^2$, $J^3$ is angular momentum along $S^2$ 
and $F_a$ is flavor charge.
This index formula is reminiscent of elliptic genus
\begin{\eq}
Z_{T^2} 
= {\rm Tr}_{\rm R}\Bigl[ (-1)^F q^{H_L} \bar{q}^{H_R }  \prod_a t_a^{F_a} \Bigr] 
= {\rm Tr}_{\rm R}\Bigl[ (-1)^F q^P  \prod_a t_a^{F_a} \Bigr] ,
\end{\eq}
which is equivalent to partition function of supersymmetric theory on $T^2$ 
with appropriate boundary conditions.
Indeed,
the work \cite{Closset:2013sxa} has shown that
the partition function on $T^2 \times S^2$ for theories only with chiral multiplets 
is exactly the same as elliptic genus of certain $\mathcal{N}=(0,2)$ theory
if we identify $J^3$ in the 4d with a flavor symmetry in the 2d.
Here we show that
this is true also for theories including vector multiplet\footnote{
Although the authors of \cite{Nishioka:2014zpa} also discussed this case,
they did not obtain final expression of the partition function on $T^2 \times S^2$. 
More concretely,
localization locus in our setup is labeled by holonomies along $T^2$ and gaugino zero modes.
Hence the partition function is given by integration over the holonomies and gaugino zero modes,
and we need to determine the integral contour.
However, their analysis ignored the gaugino zero modes and did not reach to final result.
Here we obtain final formula for the partition function by fully taking into account 
the localization locus including the gaugino zero modes.
} by using supersymmetry localization \cite{Pestun:2007rz}.
Namely we will find that
the partition function of 4d $\mathcal{N}=1$ supersymmetric gauge theory on $T^2 \times S^2$ 
is exactly the same as the one of 
certain 2d $\mathcal{N}=(0,2)$ supersymmetric gauge theory on $T^2$,
which has been recently studied well in \cite{Benini:2013xpa,Benini:2013nda,Gadde:2013dda}.
By using the recent result on the elliptic genus,
we find that
the $T^2 \times S^2$ partition function is described by Jefferey-Kirwan residue formula as in \cite{Benini:2013xpa},
\begin{\eq}
Z_{T^2 \times S^2} 
= \frac{1}{|W|} \sum_{u_\ast \in \mathcal{M}_{\rm sing}^\ast}
{\rm JKRes}_{u=u_\ast}(\mathbf{Q}(u_\ast ),\eta )\ Z_{\rm 1-loop}(\tau ,u,\sigma , \xi_a ) ,
\label{eq:main}
\end{\eq}
where we will give several definitions in next section.

Our result shows that
if we consider certain 4d supersymmetric gauge theory on $T^2 \times S^2$,
then we have corresponding 2d supersymmetric gauge theory on $T^2$, which gives the same partition function.
This fact enables us to find 
nontrivial relations between properties of 4d and 2d supersymmetric gauge theories.
For instance, we will discuss that
$(0,2)$ triality \cite{Gadde:2013lxa} in two dimensions, proposed by Gadde-Gukov-Putrov,
comes from 4d Seiberg duality \cite{Seiberg:1994pq}.
In other examples,
we identify 4d theories giving elliptic genera
of K3, M-strings and E-strings.\\
In the example of K3,
we find that
there are two 4d theories giving the elliptic genus of K3.
This would imply new four-dimensional duality.

This paper is organized as follows.
In section \ref{sec:result},
we summarize our formula on supersymmetric partition function on $T^2 \times S^2$.
In section \ref{sec:constration},
we construct 4d $\mathcal{N}=1$ supersymmetric theory on $T^2 \times S^2$.
This section is almost review of the previous works \cite{Closset:2013sxa,Closset:2013vra,Dumitrescu:2012ha}.
In section \ref{sec:localization},
we show that
the supersymmetric partition function on $T^2 \times S^2$
is equivalent to elliptic genus of certain 2d $\mathcal{N}=(0,2)$ theory.
In section \ref{sec:extended}, we discuss extended supersymmetric cases.
In section \ref{sec:examles}, we give several interesting examples.
Section \ref{sec:conclusion} is devoted to conclusion and discussions.

\subsubsection*{Note added}
When our paper was ready for submission to the arXiv,
there appeared a paper \cite{Benini:2015noa} which has some overlaps with ours.

\section{Summary of our main result}
\label{sec:result}
In this section we summarize our formula of
partition function of 4d $\mathcal{N}=1$ supersymmetric gauge theory on $T^2 \times S^2$.
If we take appropriate boundary conditions,
then this partition function is denoted as the supersymmetric index \cite{Closset:2013sxa}
\[
Z_{T^2 \times S^2} = {\rm Tr}\Bigl[ (-1)^F q^P x^{J_3} \prod_a t_a^{F_a} \Bigr] ,
\]
which we will derive in sec.~\ref{sec:index}.
The parameters $q,x$ and $t_a$ are given by
\begin{\eq}
q=e^{2\pi i\tau},\quad x=e^{2\pi i\sigma},\quad t_a = e^{2\pi i\xi_a} ,
\label{eq:fugacity}
\end{\eq}
where $(\tau ,\sigma )$ and $\xi_a$ are
complex structures of $T^2 \times S^2 $ and 
fugacity of flavor symmetry under consideration, respectively.
Then we will show that
the partition function on $T^2 \times S^2 $ is given by \eqref{eq:main}
\[
Z_{T^2 \times S^2} 
= \frac{1}{|W|} \sum_{u_\ast \in \mathcal{M}_{\rm sing}^\ast}
{\rm JKRes}_{u=u_\ast}(\mathbf{Q}(u_\ast ),\eta )\ Z_{\rm 1-loop}(\tau ,u,\sigma , \xi_a ) ,
\]
where $|W|$ is the order of the Weyl group of gauge group $G$.
Several definitions are in order.
First $Z_{\rm 1-loop}$ is roughly one-loop determinant around saddle point of localization and 
$u$ is left-moving component of a holonomy along $T^2$,
which takes values in Cartan subalgebra of the gauge group.
As we will see,
we can rewrite $Z_{\rm 1-loop}$ as
a product of one-loop determinants of 2d $\mathcal{N}=(0,2)$ multiplets on $T^2$
with charges $\mathbf{Q}(u)$ and fugacities $\sigma ,\xi_a$ of certain flavor symmetries.
$\mathcal{M}_{\rm sing}^\ast$ is
set of poles of $Z_{\rm 1-loop}$ satisfying certain conditions.
We will also explain these more precisely in next subsection.

${\rm JKRes}_{u=u_\ast}(\mathbf{Q}(u_\ast ),\eta )$
denotes a residue operation called the Jeffrey-Kirwan (JK) residue \cite{JK,Witten:1992xu},
which will be defined in sec.~\ref{sec:JK}.
The parameter $\eta$ takes values in the dual space of  Cartan subalgebra and
we have  a freedom to take $\eta$ in arbitrary nonzero values.
Although the each term in the summand depends on $\eta$,
we can show that 
the total expression is independent of choice of $\eta$.

\subsection{One-loop determinant and singularities}
\label{sec:def1loop}
Here we explain our formula for one-loop determinant in detail.
We denote $Z_{\rm 1-loop}$ as
\begin{\eq}
Z_{\rm 1-loop} (\tau ,u, \sigma ,\xi_a ) = Z_V (\tau ,u ) \prod_i Z_{\mathbf{R}_i}^{(\mathbf{r}_i )} (\tau ,u, \sigma ,\xi_a ) ,
\end{\eq}
where $Z_V$ is the contribution from 4d $\mathcal{N}=1$ vector multiplet 
while $Z_{\mathbf{R}_i}^{(\mathbf{r}_i )}$ is the one of 4d $\mathcal{N}=1$ chiral multiplet
with the representation $\mathbf{R}_i$ of $G$ and magnetic charge $\mathbf{r}_i$. 
The chiral multiplets on $T^2 \times S^2$
generally have the magnetic charge
because we need R-symmetry background gauge field with monopole configuration
in order to keep supersymmetry \cite{Closset:2013vra,Closset:2013sxa}.
We can also turn on magnetic background gauge field of flavor symmetry with integer flux $g$
as explained in sec.~\ref{sec:background}.
Hence if the chiral multiplet has R-charge $r$ and flavor charge $q_f$,
and turn on the magnetic flux $g$ of flavor symmetry,
then the magnetic charge $\mathbf{r}$ is given by
\begin{\eq}
\mathbf{r} = r+q_f g ,
\label{eq:magnetic_charge}
\end{\eq}
which should be integer in order to
satisfy quantization condition for the magnetic flux on $S^2$.
The contribution from the vector multiplet with gauge group\footnote{
For simplicity, we assume that
non-Abelian part of $G$ is simply-connected.
} $G$ is simply given by
\begin{\eq}
Z_V (\tau ,u )
= \left( \frac{2\pi\eta^2 (q)}{i}\right)^{|G|}
  \prod_{\alpha\in G} \frac{i\theta_1 (\tau |\alpha (u))}{\eta (q)} \prod_{a=1}^{|G|} du_a ,
\end{\eq}
where $|G|$ and $\alpha$ are the rank and root of the gauge group $G$, respectively.
Note that this is exactly the same as
the contribution coming from 2d $\mathcal{N}=(0,2)$ vector multiplet 
in elliptic genus formula.
The contribution from the chiral multiplet depends on the magnetic charge: 
\begin{\eq}
Z_{\mathbf{R}}^{(\mathbf{r} )} (\tau ,u, \sigma ,\xi_a )
 =\left\{ \begin{matrix}
 \prod_{m=-\frac{\mathbf{r}}{2}+1}^{\frac{\mathbf{r}}{2}-1} 
Z_{\Lambda ,\mathbf{R}} (\tau, u, m\sigma  +\sum_a q_f^{a} \xi_a ) 
& {\rm for}\ \mathbf{r}>1  \cr
1& {\rm for}\ \mathbf{r}=1 \cr
 \prod_{m=-\frac{|\mathbf{r}|}{2}}^{\frac{|\mathbf{r}|}{2}} 
Z_{\Phi  ,\mathbf{R}} (\tau, u, m\sigma  +\sum_a q_f^{a} \xi_a ) 
& {\rm for}\ \mathbf{r}<1
\end{matrix}\right. ,
\end{\eq}
where $Z_{\Lambda ,\mathbf{R}} $ and $Z_{\Phi ,\mathbf{R}} $
are the same as contributions from 2d $\mathcal{N}=(0,2)$ Fermi and chiral multiplets, respectively.
These are explicitly given by
\begin{\eq}
Z_{\Lambda ,\mathbf{R}} (\tau, u,  y ) 
=  \prod_{\rho\in \mathbf{R}} \frac{i\theta_1 (\tau | \rho (u) +y )}{\eta (q)}  ,\quad
Z_{\Phi ,\mathbf{R}} (\tau, u,  y ) 
 = \prod_{\rho\in \mathbf{R}} \frac{i\eta (q)}{\theta_1 (\tau |\rho (u)  +y)} .
\end{\eq}

When we have chiral multiplets with $\mathbf{r}>1$,
the one-loop determinant has poles.
These poles are defined as hyperplanes in $\mathbb{C}^{ |G|}$.
Denoting such hyperplane from each $Z_{\Phi ,\mathbf{R}_i}$ as $H_i$,
the hyper plane $H_i$ is given by
\begin{\eqa}
H_i = \{ \rho_i (u) +K_i (\xi )=0 & ({\rm mod}\ \mathbb{Z} +\tau\mathbb{Z}) \} ,
\end{\eqa}
where $K_i$ is weight of flavor symmetry group.
Using $H_i$, we define the singular hyperplane
\begin{\eq}
\mathcal{M}_{sing} = \cup_i H_i ,
\end{\eq}
and the set of points $u_\ast$ in $\mathcal{M}_{sing}^\ast$:
\begin{\eq}
\mathcal{M}_{sing}^\ast 
= \left\{ u_\ast \in \mathcal{M}_{sing} | 
{\rm at\ least}\ |G|\ {\rm linearly\ independent}\ H_i{\rm 's\ meet\ at}\ u_\ast \right\}
\end{\eq}
It is convenient to introduce the set $\mathbf{Q}(u_\ast )$ of charges for given $u_\ast \in \mathcal{M}_{\rm sing}^\ast$ by
\begin{\eq}
\mathbf{Q}(u_\ast ) = \{ \rho_i | u_\ast \in H_i \} .
\end{\eq}
For a technical reason,
we assume that
the set $\mathbf{Q}(u_\ast )$ is contained in a half-space of the weight space.
This condition is called projective \cite{Szenes}.
If the number of hyperplanes intersecting at $u_\ast$ is $|G|$,
then the hyperplane arrangement is always projective.
Also, even if it is not projective,
we can usually deform fugacities to reduce the number of hyperplanes at $u_\ast$
and make the arrangement projective.

\subsection{Jefferey-Kirwan residue formula}
\label{sec:JK}
Here we explain the Jefferey-Kirwan residue operation for projective hyperplane arrangement.
Suppose that $n$ linearly independent hyperplanes $H_i$'s meet at\footnote{
We can repeat similar argument for generic $u_\ast$ just by shifting the coordinates.
} $u=0$:
\begin{\eq}
H_i = \left\{ u\in \mathbb{C}^{|G|} | Q_i  (u)=0 \right\} ,
\end{\eq} 
where $i=1,\cdots ,n$.
If this hyperplane arrangement is projective,
then JK residue operation is defined by
\begin{\eq}
{\rm JKRes}_{u=0}(Q_\ast ,\eta ) 
\frac{du_1 \wedge \cdots\wedge du_{|G|}}{Q_{j_1} (u) \cdots Q_{j_{|G|}} (u)}
=\left\{ \begin{matrix} 
|{\rm det}(Q_{j_1} ,\cdots ,Q_{j_{|G|}} )|^{-1} & {\rm if}\ \eta\in {\rm Cone}(Q_{j_1} ,\cdots , Q_{j_{|G|}} ) \cr
0 & {\rm otherwise}
\end{matrix}\right. ,
\end{\eq}
where $Q_i$ is the vector, whose $a$-th component is given by coefficient of $u_a$.

For rank-1 case,
the JK residue formula is especially simpler:
\begin{\eq}
{\rm JKRes}_{u=0}(\{ q \} ,\eta ) \frac{du}{u } 
=\left\{ \begin{matrix} 
{\rm sign}(q) & {\rm if}\ \eta q >0 \cr
0                 & {\rm if}\ \eta q <0 
\end{matrix}\right. ,
\end{\eq}
Then the partition function is given by
\begin{\eq}
Z_{T^2 \times S^2}
=   \frac{1}{|W|}\sum_{u_+ \in\mathcal{M}_{\rm sing}^+} \frac{1}{2\pi i}\oint_{u=u_+} Z_{\rm 1-loop}
= -\frac{1}{|W|}\sum_{u_- \in\mathcal{M}_{\rm sing}^-} \frac{1}{2\pi i}\oint_{u=u_-} Z_{\rm 1-loop} ,
\end{\eq}
where $\mathcal{M}_{\rm sing}^\pm$ is 
points in $\mathcal{M}_{\rm sing}^\ast$ coming from charges $\pm$ of the gauge group.

\section{Four-dimensional $\mathcal{N}=1$ theory on $T^2 \times S^2$}
\label{sec:constration}
In this section,
we review\footnote{
We mainly follow a notation of \cite{Assel:2014paa}.
} a construction of 4d $\mathcal{N}=1$ supersymmetric theory on $T^2 \times S^2$ \cite{Closset:2013sxa} and
discuss some of its properties.

\subsection{$T^2 \times S^2$}
\subsubsection{Definition and complex structures}
In this paper we study supersymmetric gauge theory on $T^2 \times S^2$.
We regard this space as a quotient one of $\mathbb{C}\times S^2$.
Denoting $w$ and $z$ as the complex coordinates of  $\mathbb{C}$ and $S^2$, respectively,
we define $T^2 \times S^2$ as the following identification: 
\begin{\eq}
(w,z) \sim (w+2\pi ,e^{2\pi i\alpha}z) \sim (w+2\pi \tau ,e^{2\pi i\beta}z) ,
\label{BC}
\end{\eq}
where $\tau = \tau_1 +i\tau_2$ is the modular parameter of $T^2$ 
and $(\alpha$, $\beta )$ are real parameters with identifications $\alpha \sim \alpha +1$, $\beta \sim \beta +1$.
We also introduce 
\begin{\eq}
\sigma = \tau \alpha  -\beta .
\end{\eq}
There are two complex structure moduli $\tau$, $\sigma$, and we have the choice up to symmetries generated by 
\begin{\eqa}
&&S:~(\tau,\sigma) \mapsto \Big{(} -\frac{1}{\tau}, \frac{\sigma}{\tau} \Big{)}, \qquad 
T:~(\tau,\sigma) \mapsto (\tau+1, \sigma), \NN\\
&&U:~(\tau,\sigma) \mapsto (\tau,\sigma+\tau),\qquad\, 
V:~(\tau,\sigma) \mapsto (\tau,\sigma+1) .
\end{\eqa}  
The metric and frames on $T^2 \times S^2$ are given by\footnote{
This corresponds to take $\Omega =1 , h =0$ and $c= \frac{2}{1+|z|^2}$ in the notation of \cite{Assel:2014paa}.
}
\begin{\eq}
ds^2 = dwd\bw +\frac{4}{(1+z\bz )^2} dzd\bz ,\quad
 e^1 =dw,\quad e^2 =\frac{2}{1+|z|^2} dz .  \label{t2xs2-metric}
\end{\eq}
It is sometimes convenient to use real coordinates $(x, y)$ and $(\theta, \varphi)$:
\begin{\eq}
w = x_4 +\tau x_3 ,\quad z= \tan{\frac{\theta}{2}}e^{i(\varphi +\alpha x_4 +\beta x_3 )} .
\end{\eq}
Then the metric and frames become
\begin{\eqa}
&&ds^2 = (dx_4 +\tau_1 dx_3 )^2 +\tau_2^2 dx_3^2 +d\theta^2 
+\sin^2{\theta} (d\varphi +\alpha dx_4 +\beta dx_3 )^2 ,\NN\\
&& e^1 = dx_4 +\tau dx_3 ,\quad
e^2 =e^{i(\varphi +\alpha x_4 +\beta x_3 )} d\theta +i\sin{\theta}(d\varphi +\alpha dx_4 +\beta dx_3 ).
\end{\eqa}

\subsubsection{$T^2 \times S^2$ as a supersymmetric background}
As discussed in \cite{Festuccia:2011ws},
off-shell supersymmetric field theory with an $R$-symmetry on a curved background
can be obtained by freezing configurations of gravity multiplet in
off-shell new minimal supergravity (SUGRA) \cite{Sohnius:1981tp,Sohnius:1982fw}.
Bosonic fields in the gravity multiplet in the new-minimal off-shell SUGRA
consist of a metric and two auxiliary vector fields $A_\mu$, $V_\mu$ where  
$A_\mu$ is an $R$-symmetry gauge field, while $V_\mu$ satisfies $\nabla^\mu V_\mu =0$. 
In order to keep SUSY in fixed curved background,
we should impose 
vanishing of variation of the gravitino $(\Psi_\mu ,\tilde{\Psi}_\mu)$:
\begin{\eqa}
&& \delta \Psi_\mu
=(\nabla_\mu -iA_\mu )\zeta +iV_\mu \zeta +iV^\nu \sigma_{\mu\nu}\zeta =0 , \NN\\
&& \delta \tilde{\Psi}_\mu
=(\nabla_\mu +iA_\mu )\tzeta -iV_\mu \tzeta -iV^\nu \tsigma_{\mu\nu}\tzeta =0 , 
\label{eq:SUGRA}
\end{\eqa}
where the variation parameters $\zeta$ and $\tzeta$ have the $R$-charges as $+1$ and $-1$, respectively. 

One can show that
our background $T^2 \times S^2$ solves the Killing spinor equations \eqref{eq:SUGRA} by\footnote{
We have used a freedom in choosing $A$ to be real.
This corresponds to take $s=1$ and $\kappa =0$ in the notation of \cite{Assel:2014paa}.
} 
\begin{\eqa}
&&V= 0,\quad
A = -\frac{i}{2}\frac{\bar{z}dz-zd\bar{z}}{(1+z\bar{z})} 
  = \frac{1}{2}(1-\cos{\theta})(d\varphi +\alpha dx_4 +\beta dx_3 ) ,\NN \\
&&\zeta_\alpha = \frac{1}{\sqrt{2}}  \begin{pmatrix} 0 \cr 1\end{pmatrix} ,\quad
\tzeta^{\dalpha} = \frac{1}{\sqrt{2}} \begin{pmatrix} 1 \cr 0\end{pmatrix} . \qquad\quad
\end{\eqa}
This fact forces each field to have magnetic charge, whose value is equal to R-charge and
hence the R-charges must be integer\footnote{
As explained in \ref{sec:background},
we can also introduce background vector multiplet for global symmetry, 
which can give additional magnetic charge.
Then the R-charge is not necessary integer
because the quantization for the magnetic flux 
impose only the sum of the magnetic charges by the R-symmetry and global symmetry
to be integer.
} due to quantization condition for magnetic flux on $S^2$.

\subsection{Vector multiplet}
The Lagrangian for the vector multiplet on $T^2 \times S^2$ is 
\begin{\eq}
\calL_{\rm vec}
= {\rm Tr}\Biggl[ \frac{1}{4} \calF^{\mu\nu}\calF_{\mu\nu} -\frac{1}{2}D^2
+\frac{i}{2}\lam \sigma^\mu D_\mu \tlam +\frac{i}{2}\tlam \tsigma^\mu D_\mu \lam \Biggr] ,
\end{\eq}
where $\calF_{\mu\nu}=\del_\mu \calA_\nu -\del_\nu\calA_\mu -i[\calA_\mu,\calA_\nu]$, and    
\begin{\eq}
D_\mu
= \nabla_\mu  -i\calA_\mu  -iq_R A_\mu . 
\end{\eq}
The $R$-charges $q_R$ of the fields $(\calA_\mu,\lambda, \tlam, D)$ are $(0,1,-1,0)$. 
This action is invariant under  
\begin{\eqa}
\delta \calA_\mu &=& i\zeta\sigma_\mu\tlam +i\tzeta\tsigma_\mu \lam ,\NN\\
\delta\lam &=& \calF_{\mu\nu}\sigma^{\mu\nu}\zeta +iD\zeta ,\NN\\
\delta\tlam &=& \calF_{\mu\nu}\tsigma^{\mu\nu}\tzeta -iD\tzeta ,\NN\\
\delta D &=& -\zeta\sigma^\mu  D_\mu \tlam     +\tzeta \tsigma^\mu  D_\mu \lam ,
\end{\eqa}
where $\zeta $ and $\tzeta$ are the commuting spinors.
Note that 
$\lambda $ and $\tlam $ are independent since we are working in the 4d Euclidean signature.
Hence, 4d $\mathcal{N}=1$ SUSY requires that
$(\mathcal{A}_\mu ,D)$ are not hermitian or anti-hermitian a priori.
In order to insure the action to be real,
we take integral contour of path integral as 
\begin{\eq}
(\mathcal{A}_\mu ,D)^\dag = (\mathcal{A}_\mu ,-D) .
\end{\eq}
Namely we practically regard $\mathcal{A}_\mu (D)$ as (anti-)hermitian
after computing the SUSY variation. 

When we have a $U(1)$-part in gauge group,
we can also consider the FI-term:
\begin{\eq}
S_{\rm FI} = -i\zeta \int d^4 x \sqrt{g} D .
\end{\eq}
Note that the Lagrangian for the vector multiplet is $\delta$-exact, 
\begin{\eq}
\mathcal{L}_{\rm vec}=\mathcal{L}_{\rm vec}^{(+)} +\mathcal{L}_{\rm vec}^{(-)}  ,
\end{\eq}
where
\begin{\eqa}
&&\mathcal{L}_{\rm vec}^{(+)}
= -\delta_\zeta \left( \frac{1}{4|\zeta |^2} 
  {\rm Tr}(\delta_\zeta \lambda )^\dag \lambda \right)
= {\rm Tr} \Bigl[ \frac{1}{4}\calF_{\mu\nu}^{(+)}\calF^{(+)\mu\nu} -\frac{1}{4}D^2 
     +\frac{i}{2}\lambda\sigma^\mu D_\mu\tlam \Bigr] ,\NN\\
&&\mathcal{L}_{\rm vec}^{(-)}
= -\delta_{\tzeta} \left( \frac{1}{4|\tzeta |^2} 
  {\rm Tr}(\delta_{\tzeta} \tlam )^\dag \tlam \right)
= {\rm Tr} \Bigl[ \frac{1}{4}\calF_{\mu\nu}^{(-)}\calF^{(-)\mu\nu} -\frac{1}{4}D^2 
     +\frac{i}{2}\tlam\tsigma^\mu D_\mu\lambda \Bigr] ,
\end{\eqa}
with
\begin{\eq}
\calF_{\mu\nu}^{(\pm )} =\frac{1}{2}\left( \calF \pm \ast \calF \right)_{\mu\nu} .
\end{\eq}

\subsection{Chiral multiplet}
The Lagrangian for the chiral multiplet is 
\begin{\eqa}
\mathcal{L}_{\rm chi}
= D_\mu \tphi D^\mu \phi  +\frac{r}{2}\tphi\phi +\tphi D\phi -\tF F 
 +i\tpsi\tsigma^\mu D_\mu \psi     +i\sqrt{2}(\tphi\lambda\psi -\tpsi\tlam\phi ) ,
\end{\eqa}
where we have assigned $R$-charges $(r,r-1,r-2,-r,-r+1,-r+2)$
to $(\phi ,\psi ,F,\tphi ,\tpsi ,\tF )$.
The supersymmetric transformation is
\begin{\eqa}
\delta\phi &=& \sqrt{2}\zeta\psi ,\NN\\
\delta\psi &=& \sqrt{2}F\zeta +i\sqrt{2}(\sigma^\mu \tzeta )D_\mu \phi ,\NN\\
\delta F &=& i\sqrt{2}\tzeta\tsigma^\mu D_\mu \psi   -2i(\tzeta\tlam )\phi ,\NN\\
\delta\tphi &=& \sqrt{2}\tzeta\tpsi ,\NN\\
\delta\tpsi &=& \sqrt{2}\tF\tzeta +i\sqrt{2}(\tsigma^\mu \zeta )D_\mu \tphi ,\NN\\
\delta \tF &=& i\sqrt{2}\zeta\sigma^\mu D_\mu \tpsi       +2i\tphi (\zeta\lambda ) .
\end{\eqa} 
Again although $(\phi ,F ,\tphi ,\tilde{F} )$ are independent complex fields a priori,
we take the following integral contour  
\begin{\eq}
(\phi ,F ,\tphi ,\tF )^\dag = (\tphi ,-\tF ,\phi ,  -F ) .
\end{\eq}
Note that 
we can rewrite the Lagrangian as SUSY variation exact:
\begin{\eq}
\mathcal{L}_{\rm chi}
= \delta_\zeta \left( \frac{1}{2|\zeta |^2} 
\left[ (\delta_\zeta \psi )^\dag \psi -\tpsi (\delta_\zeta \tpsi)^\dag  \right] 
+2i\tphi \zeta^\dag \lambda \phi
\right) .  \label{eq:chi-Q-exact}
\end{\eq}

\subsection{Background vector multiplet and boundary condition}
\label{sec:background}
We can introduce background vector multiplet for global symmetries with keeping supersymmetry. 
For example, 
when we consider an Abelian flavor symmetry $U(1)_f$, 
supersymmetric configuration for the background vector multiplet 
is given by
\begin{\eqa}
v_{\mu} d x^{\mu}
&=& v_\omega d\omega +v_{\bar{\omega}}d\bar{\omega}-ig\frac{\bar{z}dz-zd\bar{z}}{2(1+z\bar{z})} \NN\\
&=& a_4 dx_4 + a_3 dx_3 +\frac{g}{2} (1- \cos \theta) (d \varphi+ \alpha dx_4 +\beta dx_3) ,
\label{eq:bgV}
\end{\eqa}
and 
\begin{\eq}
D =\frac{q_f g}{2} ,
\end{\eq}
where $(a_4, a_3)$ is the flat connection along $T^2$ direction and 
this gives the fugacity $\xi$ of the flavor symmetry introduced in \eqref{eq:fugacity} by 
\begin{\eq}
\xi = \tau a_4 -a_3 .
\end{\eq}
Then the Lagrangian for the chiral multiplet with charge $q_f$ for $U(1)_f$  slightly changes to
\begin{\eqa}
\mathcal{L}_{\rm chi}
= D_\mu \tphi D^\mu \phi  +\frac{\mathbf{r}}{2}\tphi\phi +\tphi D\phi -\tF F 
 +i\tpsi\tsigma^\mu D_\mu \psi     +i\sqrt{2}(\tphi\lambda\psi -\tpsi\tlam\phi ) ,
\end{\eqa}
where the covariant derivatives include the background gauge field $v_\mu$ and
$\mathbf{r}$ is the total magnetic charge defined by \eqref{eq:magnetic_charge},
\[
{\bf r}=r+q_f g .
\]
Because of the quantization condition for the magnetic flux,
the total magnetic charge ${\bf r}$ has to be integer.
Note that when we turn on the background magnetic gauge field of the global symmetry,
the R-charge $r$ can be non-integer depending 
on the magnetic charges coming from the other global symmetries.
Under the identification (\ref{BC}), 
every field $\Phi$ 
has the twisted boundary condition:
\begin{\eqa}
\Phi \sim e^{i \pi \bf{r} \alpha} \Phi, \quad \Phi \sim e^{i \pi \bf{r} \beta} \Phi .
\end{\eqa}

\subsection{Supersymmetry algebra and index formula}
\label{sec:index}
Denoting $\delta=\delta_\zeta + \delta_{\tzeta}$, the supersymmetry transformation generates 
\begin{\eqa}
\{ \delta_\zeta ,\delta_\zeta \}  =\{ \delta_{\tzeta} ,\delta_{\tzeta} \} =0 ,\quad
[ \delta_\zeta ,\delta_K ]  =[ \delta_{\tzeta} ,\delta_K ] =0 ,\quad
\{ \delta_\zeta ,\delta_{\tzeta} \} =2i\delta_K ,
\label{eq:algebra}
\end{\eqa}
where
\begin{\eq}
\delta_K = \calL_K -iK^\mu \calA_\mu - i K^\mu (q_R A_\mu +q_f v_\mu ). 
\end{\eq}
$\calL_K$ is a Lie derivative along the vector field $K$ given by
\begin{\eq}
K=\del_{\bw} =\frac{1}{2i\tau_2} (\tau \del_4 -\del_3 -\sigma \del_\varphi ) .
\end{\eq}  
If we identify the $x_3$-direction as ``time" circle and $x_4$-direction as ``spatial" circle,
then our partition function on $T^2 \times S^2$ can expressed as
\begin{\eq}
Z_{T^2 \times S^2} ={\rm Tr}\Bigl[ (-1)^F e^{-2\pi H} \Bigr] ,
\end{\eq}
where $H$ is Hamiltonian.
The SUSY algebra \eqref{eq:algebra} implies that
the partition function is contributed only by the states satisfying\footnote{
Here we assume discreteness of the energy spectrum.
Otherwise, the partition function would be non-holomorphic as in elliptic genera for non-compact manifolds (see e.g. \cite{Harvey:2014nha}).
}
\begin{\eq}
H = -i \left( \tau P +\sigma J_3 +(\tau a_4 -a_3 )q_f  \right) ,
\end{\eq} 
while the other states are canceled between bosonic and fermionic states.
Thus our partition function is interpreted as the index\footnote{
We have some other equivalent ways to get the same index. 
One way is to consider the identifications $(w,z)\sim (w+2\pi ,z) \sim (w+2\pi\tau ,z)$ and
impose the twisted boundary condition $\Phi (w+2\pi ,z) =x^{-J_3} \prod_a t_a^{-F_a} \Phi (w ,z)$,
or  $\Phi (x_3 +2\pi ,x_4 ,z ) =e^{-2\pi i\alpha J_3} \prod_a e^{-2\pi i a_3 F_a} \Phi (x_3 ,x_4  ,z)$, 
$\Phi (x_3  ,x_4 +2\pi ,z ) =e^{-2\pi i\beta J_3} \prod_a e^{-2\pi i a_4 F_a} \Phi (x_3 ,x_4  ,z)$.
} \eqref{eq:4dindex},
\[
Z_{T^2 \times S^2} = {\rm Tr}\Bigl[ (-1)^F q^P x^{J_3} \prod_a t_a^{F_a} \Bigr] .
\]

\subsection{Anomaly cancellation}
Since we have the non-trivial background gauge fields,
expressions for anomaly are slightly modified: 
\begin{\eq}
\del_\mu J^\mu \propto {\rm Tr}\Bigl[ \epsilon^{\mu\nu\rho\sigma} 
( \calF_{\mu\nu} +(r-1)F_{\mu\nu}^R +\sum_a q_f^{(a)}f_{\mu\nu}^{(a)} ) 
(\calF_{\rho\sigma} +(r-1)F_{\rho\sigma}^R +\sum_a q_f^{(a)}f_{\rho\sigma}^{(a)}) \Bigr] ,
\end{\eq}
where $\calF_{\mu\nu}$, $F_{\mu\nu}^R$ and $f_{\mu\nu}^{(a)}$,
are field strengths of gauge symmetry, R-symmetry and flavor symmetries, respectively.
Since the background field strengths are nontrivial only along $S^2$,
the divergence $\del_\mu J^\mu$ of the current includes the terms proportional to
\begin{\eq}
{\rm Tr}\Bigl[ \epsilon^{\mu\nu\rho\sigma}  \calF_{\mu\nu} \calF_{\rho\sigma}  \Bigr] ,\quad
 (\mathbf{r}-1) {\rm Tr}\Bigl[  \epsilon^{pq} \calF_{pq} \Bigr] .
\end{\eq}
where $\mathbf{r}=r+\sum_a q_f^{(a)}g^{(a)}$ and indices $p,q$ denote coordinates of $T^2$ direction.
The 1st one gives the standard anomaly formula in 4d without nontrivial background
while the 2nd one is particular for our setup. 
Noting the number of zero modes along $S^2$ is $\mathbf{r}-1$ for 2d positive chirality
and $1-\mathbf{r}$ for 2d negative chirality,
we can easily see that
this is the same as anomaly of 2d theory obtained by dimensional reduction along $S^2$.
Thus,
we also require the 2d gauge anomaly cancellation in addition to the standard 4d cancellation condition.

The 2d gauge anomaly cancellation condition becomes simplified 
after we impose the 4d gauge anomaly cancellation condition.
For example,
the standard 4d cancellation condition for $U(1)_R \times G\times G$ type anomaly is
\begin{\eq}
\sum_i (r^{(i)}-1)T_{\mathbf{R}_i} +T_{\rm adj.} =0 ,
\end{\eq}
while the 2d condition for $G\times G$ type anomaly is
\begin{\eq}
\sum_i (\mathbf{r}^{(i)}-1)T_{\mathbf{R}_i} +T_{\rm adj.} =0 .
\end{\eq}
If we use the 4d condition,
then the 2d condition becomes
\begin{\eq}
\sum_i \sum_a q_f^{(i,a)} g^{(a)} T_{\mathbf{R}_i} =0 ,
\end{\eq}
which is nothing but the conservation condition.

\section{Partition function on $T^2 \times S^2$ and elliptic genus}
\label{sec:localization}
In this section 
we show that the partition function on $T^2 \times S^2$ 
is the same as the elliptic genus of 2d $\mathcal{N}=(0,2)$ theory
described by zero modes along $S^2$.

\subsection{Sketch of derivation}
Our proof takes the following steps.
\begin{enumerate}
\item We apply localization method to 
the partition function of the 4d $\mathcal{N}=1$ SUSY theory on $T^2 \times S^2$.
If we denote fields on $T^2 \times S^2$ as $\Phi_{T^2 \times S^2}$,
then the SUSY localization tells us that
the partition function on $T^2 \times S^2$ is independent of $Q$-exact deformation:
\begin{\eq}
Z_{T^2 \times S^2}
= \int [D\Phi_{T^2 \times S^2}] e^{-S_{T^2 \times S^2}}
= \int [D\Phi_{T^2 \times S^2}] 
e^{-S_{T^2 \times S^2}-tQ_{T^2 \times S^2} V_{T^2 \times S^2}} ,
\end{\eq}
where $Q_{T^2 \times S^2}$ is supersymmetric transformation on $T^2 \times S^2$. 
Here we choose the fermionic functional $V_{T^2 \times S^2}$
such that $Q_{T^2 \times S^2} V_{T^2 \times S^2} $ becomes the action itself without the FI-term.
Then we can take the limit\footnote{
More precisely,
we choose $tQ_{T^2 \times S^2} V_{T^2 \times S^2} =t_v S_{\rm vec} +t_c S_{\rm chi} $ and
consider the limits $t_v \rightarrow\infty$ and $t_c \rightarrow\infty$.
However, these limits should be taken carefully in our setup contrast to usual analysis by localization. 
As we will see in next subsection, there is a problem similar to analysis of elliptic genus. 
} $t\rightarrow \infty$ and saddle point method gives exact result.

\item We consider gaussian fluctuation around the saddle point $\Phi_0$ and
perform its KK mode expansion along $S^2$.
Then we will show that
the KK mode expansion can be rewritten\footnote{
Strictly speaking, we will show this equality in a specific gauge
and we have already included gauge-fixing actions in each term.
} as
gaussian fluctuation of the action of 2d $\mathcal{N}=(0,2)$ theory on $T^2$ around $\Phi_0$:
\begin{\eq}
\left. Q_{T^2 \times S^2} V_{T^2 \times S^2} [\Phi_{T^2 \times S^2}] \right|_{\rm gaussian}
= \sum_{J=j_0}^\infty  \left. S_{T^2 }^{(J)} [ \Phi_0 ,\Phi_{T^2 }^{(J)} ] \right|_{\rm gaussian} +(\cdots ) ,
\end{\eq}
where $J$ is angular momentum along $S^2$ and $j_0$ is the one of zero modes.
The symbol ``$(\cdots )$'' denotes 
non-zero modes of vector multiplets and gauge-fixing action,
whose effect is trivial as shown below.
Noting that the supersymmetric action on $T^2$ is also $Q$-exact,
we find
\begin{\eq}
Z_{T^2 \times S^2}
=\lim_{t\rightarrow\infty} \int \prod_{J=j_0}^\infty \left( [D\Phi_{T^2 }^{(J)} ] e^{-tQ_{T^2 } V_{T^2 }^{(J)} } \right) .
\end{\eq}
Thus, 
the $T^2 \times S^2$ partition function is exactly the same as
the elliptic genus of the 2d $\mathcal{N}=(0,2)$ theory with infinite multiplets.

\item Fortunately,
we already know formula for the elliptic genus of the 2d $\mathcal{N}=(0,2)$ theory
thanks to the previous studies \cite{Benini:2013xpa,Benini:2013nda,Gadde:2013dda}.
Hence, just by using their results,
we can obtain exact expression for the $T^2 \times S^2$ partition function.
As a result,
contributions from non-zero modes along $S^2$ are trivial and we find
\begin{\eq}
Z_{T^2 \times S^2}
=\lim_{t\rightarrow\infty} \int [D\Phi_{T^2 }^{(j_0 )} ] e^{-tQ_{T^2 } V_{T^2 }^{(j_0 )} }  = Z_{T^2 } ,
\end{\eq}
which is the elliptic genus of 2d $\mathcal{N}=(0,2)$ theory described by zero modes along $S^2$.

\end{enumerate}

\subsection{Localization locus and danger of naive saddle point analysis}
We would like to compute the partition function
\begin{\eq}
Z_{T^2 \times S^2} = \int [D\Phi ]\  e^{-S_{\rm FI}-S_{\rm vec}-S_{\rm chi}}.
\end{\eq}
If we choose $S_{\rm vec}$ and $S_{\rm chi}$ 
as the deformation terms of the localization,
then we find
\begin{\eq}
Z_{T^2 \times S^2} =\lim_{t_v , t_c \rightarrow \infty} 
\int [D\Phi ]\  e^{-S_{\rm FI} -t_v S_{\rm vec} -t_c S_{\rm chi}} .
\end{\eq}
Imposing $S_{\rm vec }=0$ and $S_{\rm chi}=0$ give the saddle point conditions. 
For vector multiplet, we find
\begin{\eq}
\calF_{\mu\nu}^{(+)} =0 ,\quad \calF_{\mu\nu}^{(-)} =0 ,\quad 
D=0,\quad
\lambda\sigma^\mu D_\mu\tlam =0,\quad \tlam\tsigma^\mu D_\mu\lambda =0.
\end{\eq}
The first and second equations show that
saddle point of gauge field is flat connection.
The last two equations imply that
the saddle points of $\lambda$ and $\tlam$ are zero modes and Cartan valued.
Namely $\lambda$ and $\tlam$
are constant proportional to $\zeta$ and $\tzeta$, respectively:
\begin{\eq}
\lambda =\lambda^{(0)}\zeta ,\quad \tlam =\tlam^{(0)}\tzeta .
\end{\eq}
Saddle point for chiral multiplet is trivial:
\begin{\eq}
\phi =0,\quad \tphi =0,\quad \psi  =0,\quad \tpsi =0 ,\quad F=0,\quad \tF =0 .
\end{\eq}

In usual story of localization method,
we can exactly compute the partition function by using naive saddle point analysis.
However, in general situation, 
such naive saddle analysis sometimes would give ill-defined result and not make sense
as seen in recent studies on elliptic genus.
We discuss that this is also our case
and we need to take the limits $t_v ,t_c \rightarrow\infty$ more carefully.
For finite $t_v$ and $t_c$,
the partition function on $T^2 \times S^2$ can be written as
\begin{\eq}
Z_{T^2 \times S^2}
= \int d^{|G|}\tilde{D} \int d^{2|G|} u \ f_{t_v ,t_c} (u,\bar{u},\tilde{D})
 \exp{\Bigl[ \frac{t_v}{2}\tilde{D}^2 -\zeta \tilde{D} \Bigr]} ,
\end{\eq}
where $\tilde{D}$ and $(u,\bar{u})$ are zero modes of $D$ and $\mathcal{A}_\mu$
along $T^2 \times S^2$.
The function $f_{t_v ,t_c} (u,\bar{u},\tilde{D})$
is the result of the path integral except $\tilde{D}$ and $(u,\bar{u})$
and has a finite value in the limit $t_c \rightarrow \infty$ for arbitrary $(u,\bar{u})$ unless we take the limit $t_v\rightarrow\infty$.
If we integrate $\tilde{D}$ out,
then the partition function in the $t_c \rightarrow \infty$ limit becomes 
\begin{\eq}
Z_{T^2 \times S^2}
=  \int d^{2|G|} u\ F_{t_v}(u,\bar{u}) ,
\end{\eq}
where
\begin{\eqa}
F_{t_v}(u,\bar{u}) 
&=&\lim_{t_c\rightarrow\infty} \int d^{|G|}\tilde{D} \ f_{t_v ,t_c} (u,\bar{u},\tilde{D})
 \exp{\Bigl[ \frac{t_v}{2}\tilde{D}^2 -\zeta \tilde{D} \Bigr]} \NN\\
&=& C_{t_v} (u,\bar{u}) \int d^{2M_\ast}\phi\
  \exp{\Biggl[ -t_c \sum_{i=1}^{M_\ast} |Q_i (u) \phi_i |^2 -\frac{1}{2t_v}\left( \sum_{i=1}^{M_\ast} |\phi_i |^2 +\zeta \right)^2  \Biggr]} .
\end{\eqa}
Here $\phi_i$ is zero modes of scalar fields in chiral multiplets along $T^2$,
whose eigenvalues $Q_i (u)$ vanish as approaching $u\rightarrow u_\ast$.
The prefactor $C_{t_v} (u,\bar{u}) $ is contribution from path integral except $(u,\bar{u})$ and $\phi_i$.
We can easily see that
the integrations over $\phi_i$ for $t_v \rightarrow\infty$
diverge for $u=u_\ast$ and hence we should take this limit carefully.
Thus we decompose the integration over $(u,\bar{u})$ as
\begin{\eq}
\int_{\mathcal{M}} d^{2|G|} u = \int_{\mathcal{M}-\Delta_\epsilon} d^{2|G|} u +\int_{\Delta_\epsilon} d^{2|G|} u ,
\end{\eq}
where $\Delta_\epsilon$ is $\epsilon$-neighborhood of $\mathcal{M}_{\rm sing}$.
Then we take the limit $\epsilon\rightarrow 0$ first for finite $t_v$
and finally take the $t_v \rightarrow \infty$, namely
\begin{\eq}
Z_{T^2 \times S^2}
=\lim_{t_v\rightarrow\infty} \lim_{\epsilon\rightarrow 0} \left(  
\int_{\mathcal{M}-\Delta_\epsilon} d^{2|G|} u\ F_{t_v}(u,\bar{u}) +\int_{\Delta_\epsilon} d^{2|G|} u\ F_{t_v}(u,\bar{u}) \right) .
\label{eq:careful}
\end{\eq}
In order to perform this procedure,
we keep $t_v$ finite in terms including the zero mode of $D$.
This is equivalent to that we do not regard the zero mode of $D$ as fluctuation around the saddle point.
In subsection \ref{sec:fluctuation},
we will consider gaussian fluctuation of the action around the saddle point except the zero mode of $D$ along $S^2$ and
compute its KK-mode expansion along $S^2$.
Then we will show that 
the action can be regarded as 2d $\mathcal{N}=(0,2)$ theory on $T^2$ with infinite multiplets.

\subsection{Gauge fixing}
Here we take the gauge
\begin{\eq}
D^p \calA_p =0 ,
\end{\eq}
where $p$ denotes the $T^2$-direction.
In order to construct gauge fixing action, we introduce BRST transformation as
\begin{\eq}
Q_B A_\mu = D_\mu c_g ,\quad Q_B c_g  = -\frac{i}{2}[c_g ,c_g ],\quad Q_B \bar{c}_g =B ,\quad Q_B B =0 ,
\end{\eq}
where $c_g$ and $\bar{c}_g$ are ghosts, and B is the Nakanishi-Lautrap field.
Then gauge fixing action is given by
\begin{\eq}
\mathcal{L}_{\rm gh} 
= Q_B {\rm Tr}\Bigl[ \bar{c}_g  D^p \calA_p \Bigr]
= B\nabla^p \calA_p +\bar{c}_g D^p D_p c_g
= B\nabla^p \calA_p +\bar{c}_g D_w D_{\bw} c_g .
\end{\eq}

\subsection{Gaussian fluctuation around the saddle point}
\label{sec:fluctuation}
In this subsection,
we study quadratic fluctuation around the localization locus.
Performing KK-mode expansion along $S^2$,
we show that
the quadratic fluctuation is the same as
the one of 2d $\mathcal{N}=(0,2)$ supersymmetric theory on $T^2$ except non-zero modes of vector multiplet.

\subsubsection{Vector multiplet}
Let us expand the action around the saddle point except $D$:
\begin{\eq}
\calA_\mu \rightarrow \calA^{(0)}_\mu +\calA_\mu ,\quad
\lambda  \rightarrow \lambda^{(0)} +\lambda ,\quad \tlam  \rightarrow \tlam^{(0)} + \tlam .
\end{\eq}
Then we find the action up to the quadratic fluctuation as
\begin{\eq}
\calL_{\rm vec} |_{\rm Gauss}
= {\rm Tr}\Bigl[ \frac{1}{4} ( F_{\mu\nu}^{(0)} )^2
-\frac{1}{2}D^2
+\frac{i}{2}\lam \sigma^\mu D^{(0)}_\mu \tlam +\frac{i}{2}\tlam \tsigma^\mu D^{(0)}_\mu \lam \Bigr] ,
\end{\eq}
where $D^{(0)}_\mu$ is covariant derivative in terms of the gauge field at the saddle point and 
\begin{\eq}
F_{\mu\nu}^{(0)}= D_\mu^{(0)}\calA_\nu -D_\nu^{(0)}\calA_\mu .
\end{\eq}
Next we expand each field by monopole spherical harmonics:
\begin{\eqa}
&&\calA_i  = \sum_{\rho =1,2}\sum_{J=1}\sum_{m=-J}^J 
                 A_{Jm}^\rho  C_{i,Jm}^\rho ,\quad 
\calA_p = \sum_{J=0}\sum_{m=-J}^J A_{p,Jm} Y_{0Jm} , \quad
D = \sum_{J=0}\sum_{m=-J}^J  D_{Jm} Y_{0Jm} ,\NN\\
&& \lambda_\alpha  =\sum_{J=1}\sum_{m=-J}^J \begin{pmatrix} \beta_{Jm}Y_{2Jm} \cr -\gamma_{Jm}Y_{0Jm}\end{pmatrix}
                      +\begin{pmatrix} 0\cr -\gamma_{00} Y_{000} \end{pmatrix} ,\quad
 \tlam^{\dalpha}  =-\sum_{J=1}\sum_m \begin{pmatrix}   \tgam_{Jm}Y_{0Jm}^\dag \cr \tbeta_{Jm}Y_{2Jm}^\dag \end{pmatrix}
                     -\begin{pmatrix}   \tgam_{00} Y_{000}^\dag \cr 0  \end{pmatrix} ,\NN\\
\end{\eqa}
where $(i,j)$ and $(p,q)$ denote the $S^2$ and $T^2$ directions, respectively. 
Here $Y_{r Jm}$ is scalar monopole spherical harmonics with magnetic charge $r$ and
$C_{i,Jm}^\rho$ is usual vector spherical harmonics with polarization $\rho$ (see 
app. \ref{sec:harmonics} for detail).
By some tedious calculations, we find
\begin{\eqa}
&&\int dz d\bz \sqrt{g_{S^2}} \calL_{\rm vec} |_{\rm Gauss} \NN\\
&\simeq &  {\rm Tr}\Biggl[
   \sum_{J=0}\sum_m \left\{ -\frac{1}{8} ( D_w^{(0)}A_{\bw ,Jm} -D_{\bw}^{(0)}A_{w,Jm} )^2 
                            +\frac{J(J+1)}{2}  A_{w,Jm} A_{\bw ,Jm}  \right\} \NN\\
&& {+}   \sum_{J=1} \sum_{m}   \left\{ 
\frac{1}{2}\sum_{\rho=1,2} (D_w^{(0)} A^{\rho \dagger}_{J m} ) (D_{\bw}^{(0)} A^{\rho}_{J m} )
  +\frac{J(J+1)}{2}  A_{Jm}^{2 \dagger} A_{Jm}^2 \right\}    
 -\frac{1}{2}\sum_{J=0}\sum_m  D^{\dagger}_{Jm} D_{Jm} \NN\\
&&+\sum_{J=0}\sum_m \tgam_{Jm}  D_{\bw}  \gamma_{Jm}
   +\sum_{J=1}\sum_m
     \left(  \tbeta_{Jm} D_{w} \beta_{Jm}             
          +i\sqrt{J(J+1)} (-\tgam_{Jm}\beta_{Jm} +\tbeta_{Jm}\gamma_{Jm} ) \right)
          \Biggr] .  \NN\\
\end{\eqa}

In sec. \ref{sec:1loop},
we will show that
contribution from the non-zero modes $(J\neq 0)$ are canceled
by non-zero modes of the ghosts\footnote{
Note that we can treat this sector by the naive saddle point analysis
since this sector is not interacting with $D_{00}$ and $(\lambda^{(0)},\tlam^{(0)})$.
}.
Hence let us focus on the zero modes:
\begin{\eq}
\int dz d\bz \sqrt{g_{S^2}} \calL_{\rm vec} |_{\rm zero\ modes} 
=  {\rm Tr}\Biggl[
   \frac{1}{4} ( D_p^{(0)}A_{q,00} -D_q^{(0)}A_{p,00} )^2   
  -\frac{1}{2} D_{00}^2   +\tgam_{00}  D_{\bw}^{(0)} \gamma_{00} \Biggr] , 
\label{vectorzero}  
\end{\eq}
which is invariant under the supersymmetric transformation
\begin{\eqa}
&& \delta A_{w,00} = \sqrt{2}\tgam_{00} -\sqrt{2}\gamma_{00},\quad
\delta A_{\bw ,00}=0, \NN\\
&& \delta \gamma_{00}
=  -\frac{i}{\sqrt{2}}  (D_3^{(0)} A_{4,00} -D_4^{(0)} A_{3,0})   -\frac{i}{\sqrt{2}}D_{00}  ,\NN\\
&& \delta \tgam_{00} 
= +\frac{i}{\sqrt{2}}  (D_3^{(0)} A_{4,00} -D_4^{(0)} A_{3,00})  -\frac{i}{\sqrt{2}}D_{00} ,\NN\\
&&  \delta D_{00} 
= -\frac{i}{\sqrt{2}} D_{\bw}^{(0)} \tgam_{00}  +\frac{i}{\sqrt{2}} D_{\bw}^{(0)} \gamma_{00} .
\end{\eqa}
These are exactly the same as
the Lagrangian and SUSY transformation of 2d $\mathcal{N}=(0,2)$ super Yang-Mills theory
described in app.~\ref{app:02SYM} around the saddle point
if we identify 
\begin{\eq}
\zeta_+ =\frac{1}{\sqrt{2}},\quad \tzeta_+ =\frac{1}{\sqrt{2}},\quad
\lam_+ =-\gamma_{00},\quad  \tlam_+ =-\tgam_{00}.
\end{\eq}
The FI-term becomes
\begin{\eq}
S_{\rm FI} = -i\zeta \int dwd\bw D_{00} ,
\end{\eq}
which is also the same as the FI-term in two dimensions.
Non-zero modes are decoupled from other sectors including chiral multiplets.

\subsubsection{Chiral multiplet}
Next, we evaluate the one-loop determinant of the chiral multiplet on $T^2 \times S^2$.
 Gaussian fluctuation of the action for chiral multiplet is
\begin{\eqa}
\left. \mathcal{L}_{\rm chi} \right|_{\rm Gauss}
= D_\mu \tphi D^\mu \phi  +\frac{\br}{2}\tphi\phi +\tphi D_{00}\phi -\tF F 
 +i\tpsi\tsigma^\mu D_\mu \psi     
+i\sqrt{2}(\tphi\lambda^{(0)}\zeta\psi -\tpsi\tzeta \tlam^{(0)}\phi ) .
\end{\eqa}
As in vector multiplet,
we perform KK-mode expansion along $S^2$.
We expand the bosonic fields as
\begin{\eqa}
&& \phi =\sum_{J=|\mathbf{r}|/2}\sum_m a_{Jm}Y_{\mathbf{r},Jm},\quad  
\tphi =\sum_{J=|\mathbf{r}|/2}\sum_{m}\tildea_{Jm}Y_{\mathbf{r},Jm}^\dag ,\NN\\
&&F =\sum_{J=|\mathbf{r}-2|/2} \sum_{m}f_{Jm}Y_{\mathbf{r}-2,Jm},\quad 
\tF =\sum_{J=|\mathbf{r}-2|/2} \sum_{m}\tildef_{Jm}Y_{\mathbf{r}-2,Jm} ,
\end{\eqa}
while mode expansion for fermions depend on $\mathbf{r}$:
\begin{\eqa}
&& \psi_\alpha  
=\left\{ \begin{matrix}
\sum_{J=j_0 +1}\sum_m \begin{pmatrix} b_{Jm}Y_{\mathbf{r}Jm} \cr -c_{Jm}Y_{\mathbf{r}-2Jm}\end{pmatrix}
           +\sum_m \begin{pmatrix} 0\cr -c_{j_0 m} Y_{\mathbf{r}-2 j_0 m} \end{pmatrix}
  & {\rm for}\ \mathbf{r}>1 \cr
 \sum_{J=j_0 +1}\sum_m \begin{pmatrix} b_{Jm}Y_{\mathbf{r}Jm} \cr -c_{Jm}Y_{\mathbf{r}-2Jm}\end{pmatrix} 
  & {\rm for}\ \mathbf{r}=1 \cr
\sum_{J=j_0 +1}\sum_m \begin{pmatrix} b_{Jm}Y_{\mathbf{r}Jm} \cr -c_{Jm}Y_{\mathbf{r}-2Jm}\end{pmatrix}
           +\sum_m \begin{pmatrix} b_{j_0 m} Y_{\mathbf{r} j_0 m} \cr 0 \end{pmatrix} 
           & {\rm for}\ \mathbf{r}<1
\end{matrix} \right.  ,\NN\\
&& \tpsi^{\dalpha}  
=\left\{ \begin{matrix}
-\sum_{J=j_0 +1}\sum_m \begin{pmatrix}   \tildec_{Jm}Y_{\mathbf{r}-2Jm}^\dag \cr \tildeb_{Jm}Y_{\mathbf{r}Jm}^\dag \end{pmatrix}
                 -\sum_m \begin{pmatrix}   \tildec_{j_0 m} Y_{\mathbf{r}-2 j_0 m}^\dag \cr 0  \end{pmatrix} 
& {\rm for}\ \mathbf{r}>1  \cr
-\sum_{J=j_0 +1}\sum_m \begin{pmatrix}   \tildec_{Jm}Y_{\mathbf{r}-2 Jm}^\dag \cr \tildeb_{Jm}Y_{\mathbf{r}Jm}^\dag \end{pmatrix} 
& {\rm for}\ \mathbf{r}=1  \cr
-\sum_{J=j_0 +1}\sum_m \begin{pmatrix}   \tildec_{Jm}Y_{\mathbf{r}-2Jm}^\dag \cr \tildeb_{Jm}Y_{\mathbf{r}Jm}^\dag \end{pmatrix}
                 -\sum_m \begin{pmatrix}  0 \cr   \tildeb_{j_0 m} Y_{\mathbf{r} j_0 m}^\dag   \end{pmatrix}
& {\rm for}\ \mathbf{r}<1  \cr
\end{matrix} \right. ,
\end{\eqa}
where
\begin{\eq}
j_0 = \frac{|\mathbf{r}-1|}{2} -\frac{1}{2} .
\end{\eq}
Note that the number of the zero modes and chirality in two dimensions depend on $\mathbf{r}$.
This property is important
because the non-zero modes do not affect the final formula for the index and
only the zero-modes give nontrivial contributions as we will see.

\subsubsection*{\underline{For $\mathbf{r}>1$}}
For $\mathbf{r}>1$, the action becomes
\begin{\eqa}
\int dzd\bz \sqrt{g_{S^2}} \mathcal{L}_{\rm chi} \left. \right|_{\rm Gauss}
&=& 
 \sum_{J=|\mathbf{r}|/2} \sum_m   \tildea_{Jm} \left( -D_p^{(0)}D^{(0)p} +\lambda_{\mathbf{r}J}^2 \right) a_{Jm}   
 -\sum_{J=|\mathbf{r}-2|/2} \sum_m  \tildef_{Jm}f_{Jm}  \NN\\
&&+\sum_{J=\mathbf{r}/2} \sum_m  \tildea_{Jm} D_{00} a_{Jm}
+\sum_{J={\frac{\mathbf{r}}{2}-1}}\sum_m \tildec_{Jm}  D_{\bw}^{(0)}  c_{Jm} \NN\\
&&  
+\sum_{J={\frac{\mathbf{r}}{2}}}\sum_m    
 \Bigl(  \tildeb_{Jm} D_{w}^{(0)} b_{Jm}             
   +i\lambda_{\mathbf{r}J} (-\tildec_{Jm} b_{Jm} +\tildeb_{Jm} c_{Jm} ) \Bigr) \NN\\
&& +i\sum_{J=\mathbf{r}/2} \sum_m  
\left( \tildea_{Jm} \lambda^{(0)} b_{Jm} -\tildeb_{Jm} \tlam^{(0)} a_{Jm} \right)   ,
\end{\eqa}
where
\begin{\eq}
\lambda_{\mathbf{r} J}=\sqrt{\left( J+\frac{1}{2}\right)^2 -\frac{(\mathbf{r}-1)^2}{4}} .
\end{\eq}
We can decompose this as
\begin{\eq}
\int dzd\bz \sqrt{g_{S^2}} \mathcal{L}_{\rm chi} 
= \sum_{J=\frac{\br}{2}-1}^\infty \sum_{m=-J}^J \mathcal{L}_\Lambda^{(J,m)}
 +\sum_{J=\frac{\br}{2}}^\infty \sum_{m=-J}^J \mathcal{L}_\Phi^{(J,m)} ,
\end{\eq}
where
\begin{\eqa}
&& \mathcal{L}_\Lambda^{(J,m)}
=  
\tildec_{Jm}  D_{\bw}^{(0)} c_{J m} 
-\tildef_{Jm}f_{Jm} +\lambda_{\br J}^2 \tildea_{Jm}  a_{Jm}
   +i\lambda_{\br J} (-\tildec_{Jm} b_{Jm} +\tildeb_{Jm} c_{Jm} )
,\NN\\
&& \mathcal{L}_\Phi^{(J,m)}
= 
 D_p^{(0)} \tildea_{Jm} D^{p(0)}  a_{Jm} +\tildea_{Jm} D_{00} a_{Jm}  
+\tildeb_{J m}  D_{w}^{(0)} b_{J m} 
 +i\left( \tildea_{Jm} \lambda^{(0)} b_{Jm} -\tildeb_{Jm} \tlam^{(0)} a_{Jm} \right)   
. \NN\\
 \label{chilag1}
\end{\eqa}
These are invariant under the transformations
\begin{\eqa}
&&\delta c_{Jm} = - f_{Jm}  +i\lambda_{\br J}a_{Jm}  ,\quad
\delta \tildec_{Jm} = - \tildef_{Jm}   -i\lambda_{\br J}\tildea_{Jm}   ,\NN\\
&&\delta f_{Jm} = - D_{\bw}^{(0)}c_{Jm} +i\lambda_{\br J}b_{Jm} ,\quad
\delta \tilde{f}_{Jm} = - D_{\bw}^{(0)}\tildec_{Jm} +i \lambda_{\br J}\tildeb_{Jm}  ,
\label{eq:4dLambda} \\
&& \delta a_{Jm} = +b_{Jm}  ,\quad  \delta\tildea_{Jm} = -\tildeb_{Jm}  ,\quad
\delta b_{Jm} =  D_{\bw}^{(0)} a_{Jm} ,\quad
\delta \tildeb_{Jm} =- D_{\bw}^{(0)} \tildea_{Jm}  , 
\label{eq:4dPhi}
\end{\eqa} 
which have been derived from the 4d SUSY transformation by using the orthogonal relation of the harmonics.
We can easily see that
$\mathcal{L}_\Lambda^{(J,m)}$ and the SUSY transformation \eqref{eq:4dLambda}
are exactly the same as the ones of 2d $\mathcal{N}=(0,2)$ Fermi multiplet
with potential $E(\tilde{a}_{Jm})$ and $\tilde{E}(a_{Jm})$
if we identify
\begin{\eqa}
&&\zeta_+ =\frac{1}{\sqrt{2}},\tzeta_+ =\frac{1}{\sqrt{2}}, \quad
\psi_+ = -c_{Jm},\quad \tpsi_+ = -\tilde{c}_{Jm},\quad
G =f_{Jm},\quad  \tilde{G} =\tilde{f}_{Jm} ,\NN\\ 
&&\psi_- = -\tildeb_{Jm},\quad \tpsi_- = b_{Jm},\quad
\phi =\tilde{a}_{Jm},\quad \tphi =a_{Jm},\quad
E(\phi ) = i\lambda_{rJ}\phi ,\quad \tilde{E}(\tphi ) = -i\lambda_{rJ}\tphi .\NN\\
\label{chiralsusy1}
\end{\eqa}
Also, under this identification,
$\mathcal{L}_\Phi^{(J,m)}$ and the SUSY transformation \eqref{eq:4dPhi}
are exactly the same as the ones of 2d $\mathcal{N}=(0,2)$ chiral multiplet.

\subsubsection*{\underline{For $\mathbf{r}<1$}}
The only difference from $\mathbf{r}>1$ is the zero modes.
The result is
\begin{\eq}
\int dzd\bz \sqrt{g_{S^2}} \mathcal{L}_{\rm chi} 
= \sum_{J=-\frac{\br}{2}+1}^\infty \sum_{m=-J}^J \mathcal{L}_\Lambda^{(J,m)}
 +\sum_{J=-\frac{\br}{2}}^\infty \sum_{m=-J}^J \mathcal{L}_\Phi^{(J,m)} .
\end{\eq}
In this case, the non-trivial contribution only comes from the modes 
$(a_{J m}, \tilde{a}_{J m}, b_{Jm}, \tilde{b}_{Jm})$ with $J=j_0$.

\subsubsection*{\underline{For $\mathbf{r}=1$}}
Similarly, we find 
\begin{\eq}
\int dzd\bz \sqrt{g_{S^2}} \mathcal{L}_{\rm chi} 
= \sum_{J=\frac{1}{2}}^\infty \sum_{m=-J}^J \mathcal{L}_\Lambda^{(J,m)}
 +\sum_{J=\frac{1}{2}}^\infty \sum_{m=-J}^J \mathcal{L}_\Phi^{(J,m)} .
\end{\eq}

\subsubsection{Gauge fixing term}
Let us consider the gauge-fixing term.
By expanding
\begin{\eq}
B=\sum_{J=0} \sum_m B_{Jm}Y_{0Jm}^\dag ,\quad
c_g =\sum_{J=0} \sum_m c_{g,Jm}Y_{0Jm} ,\quad
\bar{c}_g =\sum_{J=0} \sum_m \bar{c}_{g,Jm}Y_{0Jm}^\dag ,
\end{\eq}
we find
\begin{\eq}
\int dzd\bz\sqrt{g_{S^2}}   \mathcal{L}_{\rm gh} 
= \sum_{J=0} \sum_m {\rm Tr}\Bigl[ B_{Jm}\nabla^p \caltA_{p,Jm} +\bar{c}_{g,Jm}\nabla_w \nabla_{\bw} c_{g,Jm} \Bigr]
\end{\eq}
Especially, the $J=0$ part
\begin{\eq}
\left. \int dzd\bz\sqrt{g_{S^2}}   \mathcal{L}_{\rm gh}  \right|_{J=0}
={\rm Tr} \Bigl[ B_{00}\nabla^p \caltA_{p,00} +\bar{c}_{g,00}\nabla_w \nabla_{\bw} c_{g,00} \Bigr] ,
\label{ghostzero}
\end{\eq}
is exactly the same as the gauge fixing action of 2d gauge theory with the gauge-fixing condition
\begin{\eq}
\nabla^p \caltA_{p,00} =0 ,
\end{\eq}
if we identify $B_{00}$ and $c_g ,\bar{c}_g$ 
with the Nakanishi-Lautrap field and ghosts in 2d, respectively.

\subsection{Twisted boundary condition on the torus}
In our setup, the fields $\Phi_{T^2 \times S^2}$ on $T^2 \times S^2$ with the magnetic charge $\mathbf{r}$
satisfy the twisted boundary conditions
\begin{\eqa}
&&\Phi_{T^2 \times S^2} (w+2\pi ,e^{2\pi i\alpha}z)
={e^{\pi i\mathbf{r}\alpha} }\Phi_{T^2 \times S^2} (w ,z) ,\NN\\
&&\Phi_{T^2 \times S^2} (w+2\pi \tau ,e^{2\pi i\beta}z)
={e^{\pi i\mathbf{r}\beta} }\Phi_{T^2 \times S^2} (w ,z) .
\end{\eqa} 
Since the spherical harmonics $Y_{rJm}$ satisfy
$Y_{rJm} (e^{2\pi i\alpha }z) =e^{\pi i r\alpha} e^{2\pi i m \alpha}Y_{rJm} (z) $,
the fields $\Phi_{T^2}$ in effective 2d theory on $T^2$ should satisfy
\begin{\eq}
\Phi_{T^2 } (w+2\pi ) ={ e^{-2\pi i m \alpha} }\Phi_{T^2 } (w ) ,\quad
\Phi_{T^2 } (w+2\pi \tau ) ={e^{-2\pi i m \beta} }\Phi_{T^2 } (w) .
\end{\eq} 

\subsection{One-loop determinant for non-zero modes in vector multiplet}
\label{sec:1loop}
We show that 
the non-zero modes of the vector multiplet and gauge-fixing term give trivial one-loop determinant.
First, the term including $(A_{w ,Jm} ,A_{\bw ,Jm})$ is
\begin{\eqa}
&&\int dz d\bz \sqrt{g_{S^2}} \calL_{\rm vec} |_{{\rm Gauss},(A_{w ,Jm} ,A_{\bw ,Jm})} \NN\\
&&= -\frac{1}{8}(A_{w,Jm} , A_{\bw ,Jm})
\begin{pmatrix}
D_{\bw}^2                 & -D_w D_{\bw} +2J(J+1) \cr
-D_w D_{\bw} +2J(J+1)  & D_{w}^2  
\end{pmatrix}
\begin{pmatrix} A_{w,Jm} \cr A_{\bw ,Jm}\end{pmatrix} . \nonumber \\
\end{\eqa}
Hence, the one-loop determinant from $(A_{w,Jm} ,A_{\bw ,Jm})$ is
\begin{\eq}
Z_{(A_w ,A_{\bw})} = {\rm Det}\Bigl[ D_w D_{\bw} -J(J+1)\Bigr]^{-1/2} ,
\end{\eq}
up to an overall constant.
We also find the contributions from $A^1_{Jm}$ and $A^2_{Jm}$ in a straightforward manner as
\begin{\eq}
Z_{A^1} = {\rm Det}\Bigl[ D_w D_{\bw}  \Bigr]^{-1/2} ,\quad
Z_{A^2} = {\rm Det}\Bigl[ D_w D_{\bw} -J(J+1)  \Bigr]^{-1/2} .
\end{\eq}
Similarly, noting 
\begin{\eq}
\int dz d\bz \sqrt{g_{S^2}} \calL_{\rm vec} |_{\rm Gauss,(\beta ,\gamma )} 
= (\tbeta_{Jm} ,  \tgam_{Jm} )
\begin{pmatrix}
D_w                 &  +i\sqrt{J(J+1)} \cr
-i\sqrt{J(J+1)}  & D_{\bw}  
\end{pmatrix}
\begin{pmatrix} \beta_{Jm} \cr \gamma_{Jm} \end{pmatrix} ,
\end{\eq}
we find
\begin{\eq}
Z_{(\beta ,\gamma )} = {\rm Det}\Bigl[ D_w D_{\bw} -J(J+1)\Bigr] .
\end{\eq}
Contribution from the ghosts is
\begin{\eq}
Z_{c_g } = {\rm Det}\Bigl[ D_w D_{\bw}  \Bigr] .
\end{\eq}
Thus we conclude that
the total contribution from the nonzero modes of the vector multiplet and gauge-fixing term 
is trivial\footnote{
Note that eigenvalues of $D_w$ and $D_{\bw}$
are common among all the fields having the same $m$. 
}: 
\begin{\eq}
Z_{(A_w ,A_{\bw} )}  Z_{A^1} Z_{A^2} Z_{(\beta ,\gamma )} Z_{c_g } =1 ,
\end{\eq}
up to an overall constant.

\subsection{Modular properties}
In this subsection, 
we study modular properties of the $T^2 \times S^2$ partition function
under the $SL(3; \mathbb{Z})$ transformation  
\begin{eqnarray*}
&&S:(\tau ,\sigma)\rightarrow \left( -\frac{1}{\tau} ,\frac{\sigma}{\tau} \right) , \quad T:(\tau ,\sigma)\rightarrow (\tau +1,\sigma ),\\
&&U:(\tau ,\sigma)\rightarrow (\tau ,\sigma +\tau ), \quad V:(\tau ,\sigma)\rightarrow (\tau ,\sigma +1 ). 
\end{eqnarray*}
For this purpose,
it is sufficient to study the one-loop determinant.

\subsubsection*{\underline{$S$-transformation}}
Under the S-transformation, the one-loop determinants transform as 
\begin{\eqa}
&&S:Z_V \rightarrow  (-i)^{|G|} \left( \prod_{\alpha} -i e^{\frac{\pi i}{\tau}\alpha^2 (u) } \right) Z_V ,\nonumber \\ 
&&S: Z_{\Lambda ,\mathbf{R}}(-1/\tau ,u/\tau ,y/\tau)
=\left( \prod_{\rho} -i e^{\frac{\pi i}{\tau}(\rho (u)+y)^2 } \right)  Z_{\Lambda ,\mathbf{R}}(\tau ,u ,y),\NN\\
&&S: Z_{\Phi ,\mathbf{R}}(-1/\tau ,u/\tau ,y/\tau)
=\left( \prod_{\rho} i e^{-\frac{\pi i}{\tau}(\rho (u)+y)^2 } \right)  Z_{\Phi ,\mathbf{R}}(\tau ,u ,y) .
\end{\eqa}
Then the one-loop determinant of the 4d chiral multiplet transforms as
\begin{\eqa}
&& Z_{\mathbf{R}}^{(\mathbf{r}>1)}(-1/\tau ,u/\tau ,\sigma/\tau ,\xi_a /\tau )
=\left( \prod_{m=-\frac{\mathbf{r}-1}{2}}^{\frac{\mathbf{r}-1}{2}} \prod_{\rho} -i e^{\frac{\pi i}{\tau}(\rho (u)+m\sigma +\sum_a q_f^a \xi_a )^2 } \right) 
Z_{\mathbf{R}}^{(\mathbf{r}>1)}(\tau ,u ,\sigma ,\xi_a ) ,\NN\\
&& Z_{\mathbf{R}}^{(\mathbf{r}<1)}(-1/\tau ,u/\tau ,\sigma/\tau ,\xi_a /\tau )
=\left( \prod_{m=-\frac{|\mathbf{r}|}{2}}^{\frac{|\mathbf{r}|}{2}} \prod_{\rho} i e^{-\frac{\pi i}{\tau}(\rho (u)+m\sigma +\sum_a q_f^a \xi_a )^2 } \right) 
Z_{\mathbf{R}}^{(\mathbf{r}<1)}(\tau ,u ,\sigma ,\xi_a ) .\NN\\
\end{\eqa}
We find that
the exponent of the prefactor is a linear combination
of the factors (\ref{gganomaly})-(\ref{ssanomaly})
appearing in a particular regularization of the one-loop determinant
and is related to anomalies in two dimensions (see app.~\ref{app:reg} for detail).

\subsubsection*{$T$-transformation}
The $T$-transformation acts on each one-loop determinant as
\begin{\eq}
T:Z_V \rightarrow e^{\frac{\pi i}{6}|{\rm Adj.}|} Z_V ,\quad
T:Z_{\Lambda ,\mathbf{R}} \rightarrow e^{\frac{\pi i}{6}|\mathbf{R}|} Z_{\Lambda ,\mathbf{R}},\quad
T:Z_{\Phi ,\mathbf{R}} \rightarrow e^{-\frac{\pi i}{6}|\mathbf{R}|} Z_{\Phi ,\mathbf{R}} .
\end{\eq}
Hence, the partition function transforms as
\begin{\eq}
T:Z_{T^2 \times S^2} \rightarrow e^{\frac{\pi i}{6}(|{\rm Adj.}| +\sum_i (\mathbf{r}_i -1)|\mathbf{R}_i | ) } Z_{T^2 \times S^2} .
\end{\eq}
The prefactor vanishes 
if we satisfy the 2d gauge anomaly cancellation condition for the 2d zero-mode theory.
 
\subsubsection*{$U$-transformation}
$U$-transformation properties of the one-loop determinants are
\begin{\eqa}
&&U:Z_V \rightarrow  Z_V ,\NN\\
&&U:Z_{\mathbf{R}}^{(\mathbf{r})} \rightarrow
\left\{  \begin{matrix} 
e^{-\frac{i\pi }{12}\mathbf{r}(\mathbf{r}^2 -1) \tau |\mathbf{R}| } Z_{\mathbf{R}}^{(\mathbf{r})} & {\rm for}\  \mathbf{r} >1 \cr
Z_{\mathbf{R}}^{(\mathbf{r})} & {\rm for}\  \mathbf{r} =1 \cr
e^{+\frac{i\pi }{12}|\mathbf{r}|(|\mathbf{r}| +1)(|\mathbf{r}| +2) \tau |\mathbf{R}| }Z_{\mathbf{R}}^{(\mathbf{r})} & {\rm for}\  \mathbf{r} <1 
\end{matrix} \right. .
\end{\eqa}
Although one would expect this prefactor can be rewritten in the language of anomalies,
we have not found any clear understandings.
It would be illuminating if one finds physical implications of this factor.
Presumably this would be related to gravitational anomalies.

\subsubsection*{$V$-transformation}
Each one-loop determinant is invariant under the $V$-transformation :
\begin{\eq}
V:Z_V \rightarrow  Z_V ,\quad
V:Z_{\mathbf{R}}^{(\mathbf{r})} \rightarrow Z_{\mathbf{R}}^{(\mathbf{r})} .
\end{\eq}
Thus the whole partition function is always invariant under the $V$-transformation.

\subsection{Localization by another deformation term}
If we consider another deformation term for localization,
then we can obtain another formula for the partition function,
which is apparently different from our main formula \eqref{eq:main}
but should be the same.
Here let us take the deformation term as
\begin{\eq}
QV=\mathcal{L}_{\rm vec} +\mathcal{L}_{\rm chi} +\mathcal{L}_m ,
\end{\eq}
where
\begin{\eq}
\mathcal{L}_m = \delta \Bigl[ {\rm Tr} (\zeta^\dag\lam -\tzeta^\dag\tlam ) (-iD) \Bigr] .
\end{\eq}
Then bosonic part of $\mathcal{L}_{\rm vec}$ plus $\mathcal{L}_m$ is given by
\begin{\eq}
\left. \mathcal{L}_{\rm vec} + \mathcal{L}_m \right|_{\rm bos.}
=\frac{1}{2}\mathcal{F}_{ip}^2 +\frac{1}{4}\mathcal{F}_{pq}^2
 +\frac{1}{2}(\mathcal{F}_{12} +D )^2  .
\end{\eq}
This leads us to the following condition for the saddle point 
\begin{\eq}
\calF_{12}=-D ,\quad \calF_{13}=\calF_{14}=\calF_{23}=\calF_{24}=\calF_{34}=0 .
\end{\eq}
General solution satisfying this condition and compatible with isometry of $T^2 \times S^2$
is non-Abelian version of the general background multiplet configuration \eqref{eq:bgV}.
Namely, the gauge field takes monopole configuration on $S^2$ and
has the holonomies along $T^2$.
Then, the $T^2 \times S^2$ partition function becomes\footnote{
The authors in \cite{Benini:2015noa} have claimed this formula
by an expectation from their result on $S^1 \times S^2$. 
}
\begin{\eq}
Z_{T^2 \times S^2} 
=\sum_{\mathbf{m}} \frac{e^{-S_{\rm FI}(\mathbf{m})}}{({\rm sym})} \sum_{u_\ast \in \mathcal{M}_{\rm sing}^\ast}
{\rm JKRes}_{u=u_\ast}(Q(u_\ast ),\eta )\ Z_{\rm 1-loop}^{(\mathbf{m})}(\tau ,u,\sigma , \xi_a ) ,
\label{eq:monopole}
\end{\eq}
where $\mathbf{m}=(m_1 ,\cdots ,m_{|G|})$ is the monopole charge vector and
the factor (sym) denotes the rank of Weyl group of unbroken gauge group.
Note that the FI-term is non-zero for general monopole configuration.
The one-loop determinant $Z_{\rm 1-loop}^{(\mathbf{m})}$ is given by
\begin{\eq}
Z_{\rm 1-loop}^{(\mathbf{m})} (\tau ,u, \sigma ,\xi_a ) 
= Z_V^{(\mathbf{m})} (\tau ,u ,\sigma ) 
\prod_i Z_{\mathbf{R}_i}^{(\mathbf{m},\mathbf{r_i} )} (\tau ,u, \sigma ,\xi_a ) ,
\end{\eq}
where $Z_V^{(\mathbf{m})}$ is the contribution from the vector multiplet:
\begin{\eq}
Z_V^{(\mathbf{m})} (\tau ,u,\sigma )
= \left( \frac{2\pi\eta^2 (q)}{i}\right)^{|G|}
  \prod_{\alpha\in G} \frac{i\theta_1 (\tau |\alpha (u)+|\alpha (\mathbf{m})|\sigma)}
                                       {\eta (q)} \prod_{a=1}^{|G|} du_a ,
\end{\eq}
$Z_{\mathbf{R}}^{(\mathbf{m},\mathbf{r} )}$ is the contribution from the chiral multiplet:
\begin{\eq}
Z_{\mathbf{R}}^{(\mathbf{m},\mathbf{r} )} (\tau ,u, \sigma ,\xi_a )
=\prod_{\rho \in\mathbf{R}} Z^{(\mathbf{r}+\rho (m) )} 
\bigl( \tau ,\rho (u)+\sum_a q_f^{a} \xi_a ,\sigma \bigr)
\end{\eq}
where
\begin{\eq}
Z^{(r )} (\tau ,y ,\sigma )
 =\left\{ \begin{matrix}
 \prod_{m=-\frac{{r}}{2}+1}^{\frac{{r}}{2}-1} 
\frac{i\theta_1 (\tau | y +m\sigma )}{\eta (q)}
& {\rm for}\ r>1  \cr
1& {\rm for}\ r=1 \cr
 \prod_{m=-\frac{|{r}|}{2}}^{\frac{|{r}|}{2}} 
 \prod_{\rho\in \mathbf{R}} \frac{i\eta (q)}{\theta_1 (\tau |y+m\sigma )} .
& {\rm for}\ r<1
\end{matrix}\right. .
\end{\eq}
Note that
the contribution from $\mathbf{m}=0$ in \eqref{eq:monopole} 
is exactly the same as our main formula \eqref{eq:main}.
Therefore,
if the both results \eqref{eq:main} and \eqref{eq:monopole} are correct,
then the contributions from nonzero monopole configurations must vanish:
\begin{\eq}
\sum_{\mathbf{m}\neq 0} \frac{e^{-S_{\rm FI}(\mathbf{m})}}{({\rm sym})} \sum_{u_\ast \in \mathcal{M}_{\rm sing}^\ast}
{\rm JKRes}_{u=u_\ast}(Q(u_\ast ),\eta )\ Z_{\rm 1-loop}^{(\mathbf{m})}(\tau ,u,\sigma , \xi_a ) 
=0\ ? 
\end{\eq}
Although the localization procedures by the two different deformation terms lead this equation,
we have not shown this equation by explicitly computing the final expression.
It is interesting if one can solve this puzzle.

\section{4d  indices and 2d extended supersymmetries}
\label{sec:extended}
In this section we study 4d theories, whose indices 
give 2d elliptic genera with larger supersymmetries.

\subsection{4d $\mathcal{N}=2$ partition function and 2d $\mathcal{N}=(2,2)$ elliptic genus}
\label{sec:2dN22}
\begin{table}[tbp]
\begin{center}
  \begin{tabular}{|c|c  c c | }
  \hline            &$G$                 &  $U(1)_R$ &   $U(1)_{\rm flavor}$ \\
\hline $\Phi$    & adj.                &   1           &    -1         \\  
         $Q$     &  $\mathbf{R}$     &   1/2          &   $1/2-q$        \\  
  $\tilde{Q}$   &  $\mathbf{\bar{R}}$   & 1/2            &   $1/2+q$       \\ \hline
  fugacity         &    -                     &   -               &    $z$   \\
magnetic flux &     -                &         1            &         1   \\ \hline
  \end{tabular}
\end{center}
\caption{Field content of 4d $\mathcal{N}=2$ theory,
which gives 2d $\mathcal{N}=(2,2)$ elliptic genus.}
\label{tab:N22}
\end{table}
Let us consider 4d theory having $\mathcal{N}=1$ vector multiplet $V$ and the matters listed 
in tab.~\ref{tab:N22} with the superpotential $W =\tilde{Q}\Phi Q$.
It is known that this theory has 4d $\mathcal{N}=2$ supersymmetry.
Note that we also turn on the magnetic flux of the $U(1)$ flavor symmetry with $g=1$.
Since the shifted R-charges should be integer on $T^2 \times S^2$, 
we take $q$ to be positive integer\footnote{
The value of $q$ should be chosen to consistent with gauge anomaly cancellation.
The result for negative integer $q$ is simply obtained by 
the replacement $\mathbf{R}\leftrightarrow\mathbf{\bar{R}}$. 
}.
One-loop determinants for the fields are given by
\begin{\eqa}
&&Z_V 
= \left( \frac{2\pi\eta^2 (q)}{i}\right)^{{\rm rank}G}
  \prod_{\alpha\in G} \frac{i\theta_1 (\tau |\alpha (u))}{\eta (q)} \prod_{a=1}^{{\rm rank}G} du_a ,\quad
Z_\Phi  =  \prod_{\rho\in {\rm adj.}} \frac{i\eta (q)}{\theta_1 (\tau |\rho (u) -z)} ,\NN\\
&& Z_Q  = \prod_{m=-\frac{q-1}{2}}^{\frac{q-1}{2}}  \prod_{\rho\in \mathbf{R}} \frac{i\eta (q)}{\theta_1 (\tau |\rho (u)  +(1-2q)z/2 +m\sigma )} ,\NN\\
&& Z_{\tilde{Q}}  = \prod_{m=-\frac{q-1}{2}}^{\frac{q-1}{2}}  \prod_{\rho\in \mathbf{R}} \frac{i\theta_1 (\tau | -\rho (u) +(1+2q)z/2 +m\sigma )}{\eta (q)} .
\end{\eqa}
Then we easily see that the product
\begin{\eq}
Z_V Z_\Phi
= \left( \frac{2\pi\eta^3 (q)}{\theta_1 (\tau |-z)} \right)^{{\rm rank}G}
  \prod_{\alpha\in G} \frac{\theta (\tau |\alpha (u))}{\theta (\tau |\alpha (u) -z)} ,
\end{\eq}
is the same as the one-loop determinant of 2d $\mathcal{N}=(2,2)$ vector multiplet \cite{Benini:2013xpa,Gadde:2013dda}, 
if we identify the flavor fugacity $z$ with the one of $U(1)_R$ symmetry in 2d. 
Also the other contribution
\begin{\eqa}
 Z_Q  Z_{\tilde{Q}}
= \prod_{m=-\frac{q-1}{2}}^{\frac{q-1}{2}}  \prod_{\rho\in \mathbf{R}} 
\frac{\theta_1 (\tau | \rho (u) -(1+2q)z/2 +m\sigma )}{\theta_1 (\tau |\rho (u) +(1-2q)z/2 +m\sigma )} ,
\end{\eqa}
is the same as the one-loop determinant of 2d $\mathcal{N}=(2,2)$ $q$-chiral multiplets
with the representation $\mathbf{R}$ and R-charge $(1-2q)$.
Thus 
the 4d $\mathcal{N}=2$ theory on $T^2 \times S^2$ with the appropriate background
gives the partition function of the 2d $\mathcal{N}=(2,2)$ theory on $T^2$.

\subsection{4d $\mathcal{N}=2$ partition function and 2d $\mathcal{N}=(0,4)$ elliptic genus}
\begin{table}[tbp]
\begin{center}
  \begin{tabular}{|c|c  c c | }
  \hline            &$G$                 &  $U(1)_R$ &   $U(1)_{\rm flavor}$ \\
\hline $\Phi$    & adj.                &   1           &    $1+2q$         \\  
         $Q$     &  $\mathbf{R}$     &   1/2          &   $-1/2-q$        \\  
  $\tilde{Q}$   &  $\mathbf{\bar{R}}$   & 1/2            &   $-1/2-q$       \\ \hline
  fugacity         &    -                     &   -               &    $z$   \\
magnetic flux &     -                &         1            &         1   \\ \hline
  \end{tabular}
\end{center}
\caption{Field content of 4d $\mathcal{N}=2$ theory,
which gives 2d $\mathcal{N}=(0,4)$ elliptic genus.}
\label{tab:N04}
\end{table}
Let us consider the theory with the same multiplets as in last subsection
but different flavor charges listed in tab.~\ref{tab:N04}.
We again have to take $q$ to be positive integer to satisfy the quantization condition of the magnetic flux on $S^2$.
Each one-loop determinants for each field is given by
\begin{\eqa}
&&Z_V 
= \left( \frac{2\pi\eta^2 (q)}{i}\right)^{{\rm rank}G}
  \prod_{\alpha\in G} \frac{i\theta_1 (\tau |\alpha (u))}{\eta (q)} \prod_{a=1}^{{\rm rank}G} du_a ,\NN\\
&&Z_\Phi  =\prod_{m=-q}^{q}  \prod_{\rho\in {\rm adj.}} 
\frac{i \theta_1 (\tau |\rho (u) +(1+2q)z+m\sigma )}{\eta (q)} ,\NN\\
&& Z_Q  = \prod_{m=-\frac{q}{2}}^{\frac{q}{2}}  \prod_{\rho\in \mathbf{R}} 
\frac{i\eta (q)}{\theta_1 (\tau |\rho (u)  -(1+2q)z/2 +m\sigma )} ,\NN\\
&& Z_{\tilde{Q}}  = \prod_{m=-\frac{q}{2}}^{\frac{q}{2}}  \prod_{\rho\in \mathbf{R}} 
\frac{i\eta (q)}{\theta_1 (\tau | -\rho (u) -(1+2q)z/2  +m\sigma )} .
\end{\eqa}
First let us consider the combination
\begin{\eqa}
Z_V Z_\Phi
&=& \Biggl[ \left( \frac{2\pi\eta^2 (q)}{i}\right)^{{\rm rank}G}
  \prod_{\alpha\in G} \frac{i\theta_1 (\tau |\alpha (u))}{\eta (q)} 
   \prod_{\rho\in {\rm adj.}} \frac{i \theta_1 (\tau |\rho (u) +(1+2q)z )}{\eta (q)}
\prod_{a=1}^{{\rm rank}G} du_a\Biggr] \NN\\
&&\times \Biggl[ \prod_{m=1}^{q}  \prod_{\rho\in {\rm adj.}} 
\frac{ i\theta_1 (\tau |\rho (u) +(1+2q)z+m\sigma )}{\eta (q)}
\frac{ i\theta_1 (\tau |-\rho (u) +(1+2q)z -m\sigma )}{\eta (q)}  \Biggr] . \NN\\
\end{\eqa}
We easily find that
the first factor is the same as the one-loop determinant of $(0,4)$ vector multiplet,
while the second is the one of the $(0,4)$ Fermi multiplets\footnote{
Note that the $(0,4)$ vector multiplet consists of
the $(0,2)$ vector multiplet and $(0,2)$ Fermi multiplet in adjoint representation.
Also, the $(0,4)$ Fermi (hyper) multiplet consists of
the $(0,2)$ Fermi (chiral) multiplets with representation $\mathbf{R}$ and $\bar{\mathbf{R}}$. 
}.
Also, the remaining part
\begin{\eqa}
 Z_Q  Z_{\tilde{Q}}
= \prod_{m=-\frac{q}{2}}^{\frac{q}{2}}  \prod_{\rho\in \mathbf{R}} 
\frac{i\eta (q)}{\theta_1 (\tau |\rho (u)  -(1+2q)z/2 +m\sigma )}
\frac{i\eta (q)}{\theta_1 (\tau | -\rho (u) -(1+2q)z/2  +m\sigma )} ,
\end{\eqa}
is the same as the one-loop determinant of $(0,4)$ hyper multiplets.

\subsection{4d  partition function and 2d $\mathcal{N}=(4,4)$ elliptic genus}
\begin{table}[tbp]
\begin{center}
  \begin{tabular}{|c|c  c  c c c| }
  \hline                   & $G$  & $U(1)_R$  & $U(1)_1$ & $U(1)_2$ & $U(1)_3$  \\
\hline  $\Phi_1$       & adj.   &   -1         &  1          &  0         & -1        \\  
          $\Phi_2$       & adj.   &   -1         &   1         &  0          & 1     \\  
          $\Phi_3$       & adj.   &    2          &   0         &  0         &  2  \\    \hline
           fugacity       &  -     &   -           &   $z$      &  $\xi_1$  & $\xi_2$     \\ 
      magnetic flux     &  -     &   1           &   1         &  0          & 0      \\ \hline  
\end{tabular}
\end{center}
\caption{A 4d theory,
which gives the elliptic genus of 2d $\mathcal{N}=(4,4)$ vector multiplet.}
\label{tab:N4SUSYvec}
\end{table}

First we consider the combination of 4d $\mathcal{N}=1$ multiplet which gives the one-loop determinant of 2d $\mathcal{N}=(4,4)$ vector multiplet. 
Suppose theory with $\mathcal{N}=1$ vector multiplet $V$ and three adjoint chiral multiplets $\Phi_i (i=1,2,3)$ whose charge assignments are listed in
tab.~\ref{tab:N4SUSYvec}.
The one-loop determinants of chiral multiplets  is given by
\begin{\eqa}
&&Z_{\Phi_1}  = \prod_{\rho\in {\rm adj.}}  \frac{i\eta (q)}{\theta_1 (\tau |\rho (u)  +z-\xi_2  )} , \quad 
Z_{\Phi_2}  = \prod_{\rho\in {\rm adj.}}  \frac{i\eta (q)}{\theta_1 (\tau |\rho (u)  +z+\xi_2  )} , \nonumber \\
&& Z_{\Phi_3}  =\prod_{\rho\in {\rm adj.}} 
\frac{i \theta_1 (\tau |\rho (u) +2 \xi_2)}{\eta (q)},  
\end{\eqa}
Then 
\begin{\eqa}
&&Z_V \prod_{i=1}^3 Z_{\Phi_i}  =\left( \frac{2\pi\eta^2 (q)}{i}\right)^{|G|} \prod_{\alpha \in G } {i \theta_1 (\tau | \alpha(u))}
\prod_{\rho \in {\rm adj.}} \frac{-i \theta_1 (\tau |\rho (u) +2 \xi_2)}
{\theta_1 (\tau |\rho (u)  +z-\xi_2  ) \theta_1 (\tau |\rho (u)  +z+\xi_2  )} \nonumber \\
\end{\eqa}
This agrees with the one-loop determinant of 2d $\mathcal{N}=(4,4)$ vector multiplet \cite{Harvey:2014nha}.

\begin{table}[t]
\begin{center}
  \begin{tabular}{|c|c  c  c c c| }
  \hline                   & $G$               & $U(1)_R$  & $U(1)_1$ & $U(1)_2$ & $U(1)_3$  \\
\hline  $Q$       &  $\mathbf{{R}}$       &   0         &  0          &   1        & -1        \\  
       $\tilde{Q}$ & $\mathbf{\bar{R}}$   &   0         &   0         &  -1          & -1     \\  
          $q$       &  $\mathbf{{R}}$        &    3          &   -1         &  1         &  0  \\    
  $\tilde{q}$       & $\mathbf{\bar{R}}$   &    3          &   -1         &  -1         &  0  \\  \hline
fugacity           &  -                         &   -           &   $z$      &  $\xi_1$  & $\xi_2$     \\ 
  magnetic flux &  -                          &   1           &   1         &  0          & 0      \\ \hline  
\end{tabular}
\end{center}
\caption{A 4d theory giving the elliptic genus of 2d $\mathcal{N}=(4,4)$  hyper multiplet.}
\label{tab:N4SUSYhyp}
\end{table}
Next we consider the 4d chiral multiplets, which give the one-loop determinant of 2d $\mathcal{N}=(4,4)$ hyper multiplet.
The field content of the 4d theory is listed in tab.~\ref{tab:N4SUSYhyp}.
The one-loop determinants are given by
\begin{eqnarray}
&&Z_{Q} = \prod_{\rho\in \mathbf{R}}  \frac{i\eta (q)}{\theta_1 (\tau |\rho (u)  +\xi_1-\xi_2  )}, \quad 
Z_{\tilde{Q}} = \prod_{\rho\in \mathbf{\bar{R}}}  \frac{i\eta (q)}{\theta_1 (\tau |\rho (u)  -\xi_1-\xi_2  )}, \nonumber \\ 
&&Z_{q} = \prod_{\rho\in \mathbf{R}}  \frac{i\theta_1 (\tau |\rho (u)  -z+\xi_1  )}{\eta (q)}, \quad 
 Z_{\tilde{q}} = \prod_{\rho\in \mathbf{\bar{R}}}  \frac{i\theta_1 (\tau |\rho (u)  -z-\xi_1  )}{\eta (q)}.
\end{eqnarray}
Then we obtain the one-loop determinant of $\mathcal{N}=(4,4)$ hyper multiplet.
\begin{eqnarray}
&&Z_{Q} Z_{\tilde{Q}} Z_{q} Z_{\tilde{q}}= \prod_{\rho\in \mathbf{R} }  
\frac{\theta_1 (\tau |\rho (u)  -z+\xi_1  )}{\theta_1 (\tau |\rho (u)  +\xi_1-\xi_2  )}
 \prod_{\rho\in \mathbf{\bar{R}}} \frac{\theta_1 (\tau |\rho (u)  -z-\xi_1  )}
                             {  \theta_1 (\tau |\rho (u)  -\xi_1-\xi_2  )} .
\end{eqnarray}

\section{Examples}
\label{sec:examles}
In this section we present four dimensional theories on $T^2 \times S^2$, 
whose indices have the same expressions as 
two dimensional elliptic genera of interesting examples.

\subsection{K3}
In this subsection 
we find two 4d theories on $T^2 \times S^2$ giving the elliptic genus of K3.
In other words, we show that
these two 4d theories have the same partition function via the K3 elliptic genus.
This would imply a new four dimensional duality.

\subsubsection{Two dimensional description}
\begin{table}[t]
\begin{center}
  \begin{tabular}{|c|c  c c|  }
  \hline            & $U(1)_1$ & $U(1)_2$ & $U(1)_R$  \\
\hline $P$       &   -2     &   -3         &   2            \\  
          $X_{1,2}$ &    1     &   0          &  0    \\  
         $Y_{1,2,3}$ &  0      &   1          &    0   \\ \hline
         fugacity   &   -       &  -          &    $z$\\\hline
  \end{tabular}
\end{center}
\caption{A 2d $\mathcal{N}=(2,2)$ theory giving the elliptic genus of K3.
This theory has the superpotential $W=Pf(X,Y)$, 
where $f(X,Y)$ is a homogeneous polynomial of $(X,Y)$ with degree $(2,3)$.
 }
\label{tab:K3}
\end{table}
First we briefly explain 2d SUSY gauge theories giving elliptic genus of K3. 
This subsubsection is essentially review of sec.~4.1 in \cite{Benini:2013xpa}.
Suppose 2d $\mathcal{N}=(2,2)\ U(1)_1 \times U(1)_2$ gauge theory
with the matters listed in tab.~\ref{tab:K3}.
One-loop determinant of this theory is 
\begin{\eq}
Z_{\rm 1-loop}
= \left[ \frac{2\pi\eta (q)^3}{\theta_1 (\tau |-z)} \right]^2
 \frac{\theta_1 (\tau | -2u_1 -3u_2 )}{\theta_1 (\tau |z -2u_1 -3u_2)} 
\left[ \frac{\theta_1 (\tau | -z +u_1 )}{\theta_1 (\tau |u_1 )}  \right]^2
\left[ \frac{\theta_1 (\tau |-z +u_2)}{\theta_1 (\tau |u_2 )}  \right]^3 du_1 \wedge du_2 .
\label{eq:K3_1loop_1}
\end{\eq}
Then $\mathcal{M}_{\rm sing}$ is given by the hyperplanes:
\begin{\eq}
H_P = \{ z-2u_1 -3u_2 =0 \} ,\quad H_X = \{  u_1 =0 \} ,\quad H_Y = \{ u_2 =0 \}
\quad ({\rm mod}\ \mathbb{Z}+\tau\mathbb{Z} ) .
\end{\eq}
Hence, we find that $u_\ast$ is intersections of $(H_P ,H_X )$, $(H_P ,H_Y )$ and $(H_X ,H_Y )$.
Note that the each charge covector is
\begin{\eq}
Q_P =(-2,-3),\quad Q_X = (1,0) ,\quad Q_Y = (0,1) .
\end{\eq}
If we take $\eta =(1,1)$, which is inside of ${\rm Cone}(Q_X ,Q_Y )$ but outside of ${\rm Cone}(Q_P ,Q_X )$ and ${\rm Cone}(Q_P ,Q_Y )$,
then non-zero JK residue comes only from the intersection $(H_X ,H_Y )$:
\begin{\eq}
Z_{T^2} =  \frac{1}{(2\pi i)^2} \oint _{u_1 =u_2 =0}du_1 du_2 \ Z_{\rm 1-loop}  .
\label{eq:K3_1}
\end{\eq}

\begin{table}[t]
\begin{center}
  \begin{tabular}{|c|c  c|  }
  \hline            & $U(1)$ & $U(1)_R$  \\
\hline $P$       &   -4     &      2            \\  
    $X_{1,2,3,4}$ &    1     &     0    \\ \hline 
         fugacity   &   -       &   $z$\\\hline
  \end{tabular}
\end{center}
\caption{Another 2d $\mathcal{N}=(2,2)$ theory giving the K3 elliptic genus.
This theory has the superpotential $W=Pf(X)$,
where $f(X)$ is a homogeneous polynomial of $X$ with degree $4$.
}
\label{tab:anotherK3}
\end{table}

There is another $\mathcal{N}=(2,2)$ theory giving the same elliptic genus.
This theory is $U(1)$ gauge theory with the matter listed in tab.~\ref{tab:anotherK3}.
By taking $\eta >0$, we find
\begin{\eq}
Z_{T^2} = \frac{\eta (q)^3}{i\theta_1 (\tau | -z)}\
\oint_{u=0} du \frac{\theta_1 (\tau | -4u ) }{\theta_1 (\tau | z-4u )}
\left( \frac{\theta_1 (\tau | -z+u )}{\theta_1 (\tau | u )} \right)^4 .
\label{eq:K3_2}
\end{\eq}
It is known that the K3 elliptic genus in standard form is \cite{Eguchi:1988vra}
\begin{\eq}
Z_{T^2} = 8\Biggl[
\left( \frac{\theta_1 (\tau |z+\frac{1}{2})}{\theta_1 (\tau |\frac{1}{2})} \right)^2
+\left( e^{\pi iz}\frac{\theta_1 (\tau |z+\frac{1+\tau}{2})}{\theta_1 (\tau |\frac{1+\tau}{2})} \right)^2
+\left( e^{\pi iz}\frac{\theta_1 (\tau |z+\frac{\tau}{2})}{\theta_1 (\tau |\frac{\tau}{2})} \right)^2
\Biggr] .
\end{\eq}
One can easily check that the expressions \eqref{eq:K3_1} and \eqref{eq:K3_2}
are the same as this standard form.

\subsubsection{Four dimensional description}
\begin{table}[t]
\begin{center}
  \begin{tabular}{|c|c  c c c c  c c|  }
  \hline            & $U(1)_1$ & $U(1)_2$ & $U(1)_R$ & $U(1)_f$  & $U(1)_P$&$U(1)_X$ & $U(1)_Y$ \\
\hline  $\Phi_{1,2}$   &    0        &     0       &    1    &   -1        &     0       &   0        &  0  \\
           $P'$       &   -2      &   -3         &   1/2      &   -1/2   &     1        & 0          &  0\\  
      $\tilde{P}'$   &   +2      &  +3         &   1/2      &   3/2     &    -1       &  0          &0 \\  
          $X_{1,2}'$ &    +1     &   0          &  1/2        &   -1/2   &     0       &   1          &  0\\  
     $\tilde{X}_{1,2}'$ &  -1   &   0          &  1/2   &    3/2         &    0       &  -1         &  0\\  
         $Y_{1,2,3}'$ &    0      &   +1          &    1/2     &    -1/2  &   0       &   0          &  1\\ 
   $\tilde{Y}_{1,2,3}'$ & 0      &   -1        &    1/2   &     3/2      &   0       &   0         &-1\\  \hline
         fugacity   &   -       &  -          &     -        &     z          & $\xi_P$  &  $\xi_X$  &$\xi_Y$  \\  
 magnetic flux   &     -       &  -        &     1         &    1         &   0         &  0         &0 \\\hline
  \end{tabular}
\end{center}
\caption{A 4d theory giving the elliptic genus of K3.
This theory has the same one-loop determinant as the 2d theory described in tab.~\ref{tab:K3}
for special fugacities.}
\label{tab:4dK3}
\end{table}
Next we find two 4d theories on $T^2 \times S^2$,
whose partition functions are the same as the K3 elliptic genus.
Let us consider 4d $U(1)_1 \times U(1)_2$ gauge theory 
with matters listed in tab.~\ref{tab:4dK3} and the superpotential
\begin{\eq}
W = \sum_{i=1}^2\tilde{P}' \Phi_i P 
+\sum_{i=1}^3\tilde{X}_i' \Phi_1 X_i' +\sum_{i=1}^2 \tilde{Y}_i' \Phi_2 Y_i' .
\end{\eq}
The one-loop determinant is given by
\begin{\eqa}
Z_{\rm 1-loop}
&=& \Biggl[ \frac{2\pi\eta (q)^3}{\theta_1 (\tau |-z)} \Biggr]^2
 \frac{\theta_1 (\tau |-3z/2 -2u_1 -3u_2 +\xi_P )}{\theta_1 (\tau |-z/2 -2u_1 -3u_2 +\xi_P )}  \NN\\
&&\Biggl[ \frac{\theta_1 (\tau | -3z/2+u_1 +\xi_X )}{\theta_1 (\tau |-z/2 + u_1 +\xi_X )}  \Biggr]^2
    \Biggl[ \frac{\theta_1 (\tau | -3z/2 +u_2 +\xi_Y )}{\theta_1 (\tau | -z/2 +u_2 +\xi_Y )}  \Biggr]^3 du_1 \wedge du_2 .
\end{\eqa}
If we take $\xi_P =3z/2 ,\xi_X =z/2 ,Y=z/2$, 
then this becomes the same as  the one-loop determinant \eqref{eq:K3_1loop_1}
of the theory in tab.~\ref{tab:K3}
and hence gives the K3 elliptic genus.

\begin{table}[tbp]
\begin{center}
  \begin{tabular}{|c|c   c c c  c |  }
  \hline            & $U(1)$  & $U(1)_R$ & $U(1)_f$  & $U(1)_P$&$U(1)_X$ \\
\hline  $\Phi$   &  0         &    1         &    -1        &     0       &   0          \\
           $P'$       &   -4       &   1/2      &   -1/2   &     1        & 0         \\  
      $\tilde{P}'$   &   +4      &   1/2      &   3/2     &    -1       &  0         \\  
          $X_{1,2,3,4}'$ &    +1  &  1/2        &   -1/2   &     0       &   1       \\  
 $\tilde{X}_{1,2,3,4}'$ &  -1   &  1/2   &    3/2         &    0       &  -1        \\  \hline
         fugacity   &   -      &     -        &    $z$          & $\xi_P$  &  $\xi_X$   \\  
 magnetic flux   &     -    &     1         &    1         &   0         &  0         \\\hline
  \end{tabular}
\end{center}
\caption{Another 4d theory giving the elliptic genus of K3.
This theory has the same one-loop determinant as the 2d theory explained in tab.~\ref{tab:anotherK3}
for special fugacities. }
\label{tab:another4dK3}
\end{table}
Let us consider another 4d $U(1)$ gauge theory,
whose matters are listed in tab.~\ref{tab:another4dK3} and 
superpotential is given by
\begin{\eq}
W = \tilde{P}' \Phi P   +\sum_{i=1}^4 \tilde{X}_i' \Phi X_i' .
\end{\eq}
One-loop determinant is given by
\begin{\eq}
Z_{\rm 1-loop} 
= \frac{2\pi \eta (q)^3}{\theta_1 (\tau | -z)} \frac{\theta_1 (\tau | -3z/2 -4u +\xi_P ) }{\theta_1 (\tau | -z/2 -4u  +\xi_P )}
\left( \frac{\theta_1 (\tau | -3z/2 +u +\xi_X  )}{\theta_1 (\tau | -z/2 + u +\xi_X )} \right)^4 .
\end{\eq}
Taking $\xi_P =3z/2$ and $\xi_X  =z/2$,
this becomes the integrand of \eqref{eq:K3_2} 
and leads also the K3 elliptic genus.

Thus we find that
the two 4d theories have the same partition function.
This implies that there is a new type of duality between the two theories.
It is interesting if we further test this relation in other observables or find any physical reasons for that.

\subsection{Elliptic genus of E-strings from 4d index}
E-strings \cite{Klemm:1996hh} are
M2-branes suspended between M5 and M9 branes in M-theory description.
It is discussed in \cite{Kim:2014dza} that
low-energy dynamics of $N$ E-strings is described by
the 2d $\mathcal{N}=(0,4)\ O(N)$ gauge theory with the field content
\begin{itemize}
\item $\mathcal{N}=(0,4)$ vector multiplet
\item $\mathcal{N}=(0,4)$ hyper multiplet in symmetric representation
\item Four $\mathcal{N}=(0,4)$ Fermi multiplets in fundamental representation ,
\end{itemize}
at its IR fixed point.
Let us consider the elliptic genus of this theory defined by
\begin{\eq}
Z =
{\rm Tr}_{\rm RR} \Biggl[ (-1)^F q^{H_L}\bar{q}^{H_R} e^{2\pi i\epsilon_1 (J_1 +J_I )}
  e^{2\pi i\epsilon_2 (J_2 +J_I )} \prod_{\ell =1}^8 e^{2\pi im_\ell F_\ell}\Biggr] .
\end{\eq}
Here $J_{1,2}$ is Cartan of $SO(4)=SU(2)\times SU(2)$
associated with rotational symmetry in four-directions,
where NS5 and D8-O8 spread out.
$F_{\ell}$ is flavor symmetry of the eight $(0,2)$ Fermi multiplets,
or equivalently Cartan of $SO(16)$ symmetry.
One-loop determinant of the $(0,4)$ theory is given by
\begin{\eqa}
&&Z_V = \left( \frac{2\pi\eta^2 (q)}{i}\right)^{|O(N)|} \prod_{\alpha \in \rm root}
\frac{i\theta_1 (\tau | \alpha (u))}{\eta (q)} 
 \prod_{\rho \in \rm anti-sym}
\frac{i\theta_1 (\tau | \epsilon_1 +\epsilon_2 +\rho (u))}{\eta (q)}  ,\NN\\
&&Z_{\rm hyper}
= \prod_{\rho\in \rm sym}\frac{i\eta (q)}{\theta_1 (\tau | \epsilon_1 +\rho (u))}
                                          \frac{i\eta (q)}{\theta_1 (\tau | \epsilon_2 +\rho (u))} ,\quad
Z_{\rm Fermi}
=\prod_{\ell =1}^8 
\prod_{\rho\in \rm fund}    \frac{i\theta_1 (\tau | m_\ell  +\rho (u))}{\eta (q)} .\NN\\
\end{\eqa}

\begin{table}[tbp]
\begin{center}
  \begin{tabular}{|c|c  c c|  }
  \hline                       & $O(N)$     & $U(1)_R$ & $SO (8)$  \\
\hline $\Phi$            & anti-sym.   &   2         &  1 \\  
          $S$                  & sym.          &  -1       &  1    \\  
         $Q_{i=1,\cdots ,8}$ & fund.     &   2         &   fund.   \\ \hline
         fugacity              &   -       &  -          &    $m_\ell$ \\
        magnetic flux         &   -       &  1          &   -  \\\hline
  \end{tabular}
\end{center}
\caption{A 4d theory giving elliptic genus of $N$ E-strings.}
\label{tab:Estring}
\end{table}

We can engineer 4d theory on $T^2 \times S^2$ giving the elliptic genus of the E-strings.
Suppose 4d $O(n)$ gauge theory with the field content summarized in tab.~\ref{tab:Estring}.
Then one-loop determinant  is given by
\begin{\eqa}
&& Z_V = \left( \frac{2\pi\eta^2 (q)}{i}\right)^{|O(N)|} \prod_{\alpha \in \rm root}
\frac{i\theta_1 (\tau | \alpha (u))}{\eta (q)} ,\quad
Z_\Phi
= \prod_{\rho \in \rm anti-sym}
\frac{i\theta_1 (\tau | \rho (u))}{\eta (q)}  ,\NN\\
&&Z_{\rm sym}
= \prod_{m=-1/2}^{1/2} \prod_{\rho\in \rm sym}\frac{i\eta (q)}{\theta_1 (\tau | k\sigma +\rho (u))} ,\quad
Z_{\rm Fermi}
=\prod_{\ell =1}^8 \prod_{\rho\in \rm fund}    \frac{i\theta_1 (\tau | m_\ell  +\rho (u))}{\eta (q)} .                                   
\end{\eqa}
Comparing the one-loop determinants,
we easily see that 
the partition function of this theory is the same as the elliptic genus of the E-strings with
\begin{\eq}
\epsilon_1 = -\epsilon_2 = \frac{1}{2} \sigma .
\end{\eq}
It is interesting if we find any physical implications of this correspondence.
For example, in F-theory setup,
the E-strings arise by wrapping D3-branes on $\mathbb{P}^1$
and hence the elliptic genus of the $N$ E-strings would be described by 
the partition function $\mathcal{N}=4$ $U(N)$ SYM on $T^2 \times \mathbb{P}^1$.
We expect that this point of view would give some insights on this correspondence. 

\subsection{M-strings}
M-strings \cite{Haghighat:2013gba} are M2-branes suspended between parallel adjacent M5-branes.
It is expected that
the low-energy dynamics of $N$-tuple M-strings is described \cite{Haghighat:2013gba} (see also \cite{Hosomichi:2014rqa}) by 
the 2d $\mathcal{N}=(0,4)$ $U(N)$ gauge theory with 
\begin{itemize}
\item  $\mathcal{N}=(0,4)$ $U(N)$ vector multiplet  
\item   $\mathcal{N}=(0,4)$  hyper multiplet in adjoint representation
\item  $\mathcal{N}=(0,4)$  hyper multiplet in fundamental representation
\item   $\mathcal{N}=(0,4)$  Fermi multiplet in fundamental representation ,
\end{itemize}
at its IR fixed point.
Here we consider the elliptic genus for the M-strings defined by
\begin{\eqa}
\mathrm{Tr} \left[ (-1)^F q^{H_L} \bar{q}^{H_R} e^{2\pi i  \epsilon_1 ( J_1 +J_2+J_4) } e^{2\pi i  \epsilon_2 ( -J_1 +J_2+J_4) }
e^{2\pi i  m J_3 }  \right] ,
\end{\eqa}
where $J_1$ $(J_2,J_3,J_4)$ is the Cartan part of generator of $SU(2)$ flavor symmetry ($SU(2)^3$ R-symmetry), respectively.  
The one-loop determinants  of $\mathcal{N}=(0,4)$ multiplets are
\begin{\eqa}
Z_{\rm vec.}&=& \left( \frac{2\pi\eta^2 (q)}{i}\right)^N \prod_{i \neq j }
\frac{i\theta_1 (\tau | u_i -u_j)}{\eta (q)} 
 \prod_{i,j=1}^{N}
\frac{i\theta_1 (\tau | \epsilon_1 +\epsilon_2 +u_i-u_j)}{\eta (q)}  ,\NN\\
Z_{\rm adj-hyp} 
&=&  \prod_{i,j=1}^N \frac{i\eta (q)}{\theta_1 (\tau | \epsilon_1 +u_i -u_j)}
                                          \frac{i\eta (q)}{\theta_1 (\tau | \epsilon_2 +u_i-u_j)} ,\NN\\
Z_{\rm fund-hyp}
&=&  \prod_{i=1}^N \frac{i\eta (q)}{\theta_1 (\tau | -\frac{\epsilon_1+\epsilon_2}{2} +u_i)}
                                          \frac{i\eta (q)}{\theta_1 (\tau | -\frac{\epsilon_1+\epsilon_2}{2} -u_i)} ,\NN\\
Z_{\rm Fermi}
&=& \prod_{i=1}^N    \frac{i\theta_1 (\tau | -m  +u_i)}{\eta (q)}  \frac{i\theta_1 (\tau | -m  -u_i)}{\eta (q)} .
\label{Mstring_2d}
\end{\eqa}

\begin{table}[tbp]
\begin{center}
  \begin{tabular}{|c|c  c c|  }
  \hline                       & $U(N)$     & $U(1)_R$ & $U(1)_f$  \\
\hline $\Phi_V$            & adj,              &   2         &  0 \\  
         $\Phi_C$            & adj,              &   -1         &  0 \\  
         $Q$                   & fund.      &   0         &   0  \\ 
         $\tilde{Q}$           & a-fund.  &   0         &   0  \\ 
         $Q'$                   & fund.     &   2          &   -1  \\ 
         $\tilde{Q}'$         & a-fund.     &   2        &   -1  \\ \hline
         fugacity              &   -       &  -          &    $m$ \\
        magnetic flux         &   -       &  1          &   0  \\\hline
  \end{tabular}
\end{center}
\caption{A 4d theory giving elliptic genus of $N$ M-strings.}
\label{tab:Mstring}
\end{table}

We can identify 4d theory giving the elliptic genus of this 2d theory.
Let us consider 4d $U(N)$ supersymmetric gauge theory with the matters listed 
in tab.~\ref{tab:Mstring}.
Then one-loop determinant is given by
\begin{\eqa}
&&Z_V = \left( \frac{2\pi\eta^2 (q)}{i}\right)^N \prod_{i \neq j}
\frac{i\theta_1 (\tau | u_i -u_j)}{\eta (q)} ,\quad
Z_{\Phi_V}
=\prod_{i,j=1}^N
\frac{i\theta_1 (\tau | u_i-u_j)}{\eta (q)}  ,\NN\\
&&Z_{\Phi_C}
= \prod_{m=-1/2}^{1/2}
\prod_{i,j=1}^N \frac{i\eta (q)}{\theta_1 (\tau | m\sigma +u_i-u_j)} ,\quad
Z_{Q} Z_{Q'}
= \prod_{i=1}^N \frac{i\eta (q)}{\theta_1 (\tau | u_i)} \frac{i\eta (q)}{\theta_1 (\tau | -u_i)},\NN\\
&&Z_{\tilde{Q}} Z_{\tilde{Q}'}
= \prod_{i=1}^N  \frac{i\theta_1 (\tau | \xi  +u_i)}{\eta (q)}
 \frac{i\theta_1 (\tau | \xi  -u_i)}{\eta (q)}  .                              
\end{\eqa}
Comparing these with \eqref{Mstring_2d},
we find that 
this is the same as the elliptic genus of the M-strings 
if we make the identification
\begin{\eq}
\epsilon_1 = -\epsilon_2 = \frac{1}{2} \sigma . 
\end{\eq}
It is attractive if we find any physical implications of this correspondence.

\subsection{4d Seiberg duality and 2d $(0,2)$ triality}
We discuss that   
4d Seiberg duality \cite{Seiberg:1994pq} for $T^2 \times S^2$ partition function
gives 2d $(0,2)$ triality \cite{Gadde:2013lxa} for elliptic genus\footnote{
Basic idea of this subsection was presented by Yuji Tachikawa
in his intensive lecture at Osaka University (in October, 2014 ).  
Handwritten notes in Japanese are available at 
\url{http://member.ipmu.jp/yuji.tachikawa/lectures/2014-osaka-kyoto/}.
We are grateful to him for this point.
}.

\subsubsection{2d $(0,2)$ triality}
\begin{table}[tbp]
\begin{center}
  \begin{tabular}{|c|c|c  c c c c |}
  \hline &chiral/Fermi                & $U(N_c)$  & $SU(N_1 )$ & $SU(N_2 )$ & $SU(N_3 )$ & $SU(2)$ \\
\hline $P_{\alpha=1,\cdots ,N_1}$&chiral       & fund    & a-fund      &   1             &    1          & 1 \\  
         $\Phi_{\beta =1,\cdots ,N_2}$&chiral & a-fund  & 1             &   fund            &    1        &1 \\  
    $\Psi_{\gamma =1,\cdots ,N_3}$ &Fermi & fund    & 1             &   1               &   fund         &1\\  
     $\Gamma_{\alpha\beta}$     &Fermi     & 1    & fund             &  a-fund      &       1     &1 \\  
       $\Omega_{s=1,2}$          &Fermi  & det &1          &    1                &    1               &fund  \\ \hline
         fugacities         &   -        &-     &    $\xi_\alpha$ &  $\eta_\beta$ & $\zeta_\gamma$  &   $\lambda_s$        \\ \hline
  \end{tabular}
\end{center}
\caption{Matter content for the 2d $\mathcal{N}=(0,2)$ $U(N_c )$ SQCD with $N_c  =(N_1 +N_2 -N_3 )/2$.
This theory has the superpotential $\Phi \Gamma P$.}
\label{tab:2dSQCD}
\end{table}
First we briefly introduce the $(0,2)$ triality proposed by Gadde-Gukov-Putrov \cite{Gadde:2013lxa}.
Let us consider the 2d $\mathcal{N}=(0,2)$ $U(N_c )$ SQCD with 
\begin{\eq}
N_c = \frac{N_1 +N_2 -N_3}{2} ,
\end{\eq}
whose matter content is summarized in tab.~\ref{tab:2dSQCD}. 
The fields $\Gamma$ and $\Omega$ are required to cancel gauge anomalies.
Then the authors in \cite{Gadde:2013lxa} have conjectured the $(0,2)$ triality,
which states that
the SQCD at an infrared fixed point is invariant under the replacements
\begin{\eq}
(N_1 ,N_2 ,N_3 )\rightarrow  (N_2 ,N_3 ,N_1 ) \rightarrow  (N_3 ,N_1 ,N_2 ) . 
\end{\eq}
This conjecture has been checked 
for some observables \cite{Gadde:2013lxa,Gadde:2014ppa,Guo:2015gha}.

Let us explicitly check the $(0,2)$ triality for the elliptic genus as in \cite{Gadde:2013lxa}.
First, one-loop determinant from each field is
\begin{\eqa}
&& Z_V = \left( \frac{2\pi\eta^2 (q)}{i}\right)^{N_c} \prod_{i\neq j} \frac{i\theta_1 (\tau | u_i -u_j )}{\eta (q)} ,\quad
Z_P =\prod_{i=1}^{N_c} \prod_{\alpha =1}^{N_1} \frac{i\eta (q)}{\theta_1 (\tau |u_i -\xi_\alpha )}, \NN\\
&&Z_\Phi =\prod_{i=1}^{N_c} \prod_{\beta =1}^{N_2} \frac{i\eta (q)}{\theta_1 (\tau | -u_i +\eta_\beta )},\quad
Z_\Psi =\prod_{i=1}^{N_c} \prod_{\gamma =1}^{N_3} \frac{i\theta_1 (\tau | u_i +\zeta_\gamma )}{\eta (q)} ,\NN\\
&&Z_\Gamma =\prod_{\alpha =1}^{N_1} \prod_{\beta  =1}^{N_2} \frac{i\theta_1 (\tau | \xi_\alpha -\eta_\beta )}{\eta (q)} ,\quad
Z_\Omega =\prod_{s =1}^2   \frac{i\theta_1 (\tau | \sum_i u_i +\lambda_s )}{\eta (q)} ,
\end{\eqa}
where the fugacities satisfy $\sum_{\alpha =1}^{N_1} \xi_\alpha =0$, $\sum_{\beta =1}^{N_2} \eta_\beta =0$,
$\sum_{\gamma =1}^{N_3}\zeta_\gamma =0$ and $\sum_{s=1}^2 \lambda_s =0$.
Hence, $\mathcal{M}_{\rm sing}$ is given by
\begin{\eq}
H_{i\alpha}^P =\{ u_i =\xi_\alpha \} ,\quad H_{i\beta}^\Phi =\{ -u_i = -\eta_\beta \} .
\end{\eq}
If we take $\eta =(1,\cdots ,1)$, then we have contributions only from $H_{i\alpha}^P$ and get
\begin{\eqa}
Z_{T^2} 
&=& \sum_{\mathcal{I}\in C(N_c , N_1 )}
\Biggl[ \prod_{\alpha^\prime \in \mathcal{I}} \prod_{\alpha \neq \mathcal{I}} \frac{i\eta (q)}{\theta_1 (\tau |\xi_{\alpha^\prime} -\xi_\alpha )} \Biggr]
\Biggl[ \prod_{\alpha^\prime \neq \mathcal{I}} \prod_{\beta =1}^{N_2} \frac{i\theta_1 (\tau | -\xi_{\alpha^\prime} +\eta_\beta )}{\eta (q)} \Biggr] \NN\\
&&\times \Biggl[ \prod_{\alpha^\prime \in \mathcal{I}} \prod_{\gamma =1}^{N_3} 
\frac{i\theta_1 (\tau | \xi_{\alpha^\prime} +\zeta_\gamma )}{\eta (q)} \Biggr]
\Biggl[ \prod_{s =1}^2   \frac{i\theta_1 (\tau | \sum_{\alpha^\prime \in\mathcal{I}} \xi_{\alpha^\prime} +\lambda_s )}{\eta (q)}  \Biggr] .
\label{eq:2dSQCD1}
\end{\eqa}
Also, taking $\eta =(-1,\cdots ,-1)$ and picking up contributions from $H_{i\beta}^\Phi$ lead us to 
\begin{\eqa}
Z_{T^2} 
&=& \sum_{\mathcal{I}\in C(N_c , N_2 )}
\Biggl[ \prod_{\beta^\prime \in \mathcal{I}} \prod_{\beta \neq \mathcal{I}} \frac{i\eta (q)}{\theta_1 (\tau |-\eta_{\beta^\prime} +\eta_\beta )} \Biggr]
\Biggl[ \prod_{\beta^\prime \neq \mathcal{I}} \prod_{\alpha =1}^{N_1} \frac{i\theta_1 (\tau | \eta_{\beta^\prime} -\xi_\alpha )}{\eta (q)} \Biggr] \NN\\
&&\times \Biggl[ \prod_{\beta^\prime \in \mathcal{I}} \prod_{\gamma =1}^{N_3} 
\frac{i\theta_1 (\tau | \eta_{\beta^\prime} +\zeta_\gamma )}{\eta (q)} \Biggr]
\Biggl[ \prod_{s =1}^2  \frac{i\theta_1 (\tau | \sum_{\beta^\prime \in\mathcal{I}} \eta_{\beta^\prime} +\lambda_s )}{\eta (q)}  \Biggr] .
\end{\eqa}
Let us introduce $\tilde{\mathcal{I}}\in C(N_1 -N_c ,N_1 )$.
Then, 
by using $\prod_{\alpha^\prime \in \mathcal{I}} =\prod_{\alpha^\prime \neq \tilde{\mathcal{I}}}$, 
we find
\begin{\eqa}
Z_{T^2} 
&=& \sum_{\tilde{\mathcal{I}}\in C(\frac{N_3 +N_1 -N_2}{2} , N_1 )}
\Biggl[ \prod_{\alpha \in \tilde{\mathcal{I}}} \prod_{\alpha^\prime \neq \tilde{\mathcal{I}}} 
\frac{i\eta (q)}{\theta_1 (\tau | -\xi_\alpha +\xi_{\alpha^\prime} )}  \Biggr]
\Biggl[ \prod_{\alpha^\prime \neq \tilde{\mathcal{I}}} \prod_{\gamma =1}^{N_3}  
\frac{i\theta_1 (\tau | \xi_{\alpha^\prime} +\zeta_\gamma )}{\eta (q)}  \Biggr] \NN\\
&&\times \Biggl[ \prod_{\alpha^\prime \in \tilde{\mathcal{I}}} \prod_{\beta =1}^{N_2} 
\frac{i\theta_1 (\tau | -\xi_{\alpha^\prime} +\eta_\beta )}{\eta (q)}  \Biggr]
\Biggl[ \prod_{s =1}^2   \frac{i\theta_1 (\tau | -\sum_{\alpha^\prime \in\tilde{\mathcal{I}}} \xi_{\alpha^\prime} +\lambda_s )}{\eta (q)}  \Biggr] . 
\label{eq:2dSQCD2}
\end{\eqa}
Comparing this with \eqref{eq:2dSQCD1},
we find that the elliptic genus is invariant under the replacements $(N_1 ,N_2 ,N_3 )\rightarrow (N_3 ,N_1 ,N_2 )$ and
$(\xi_\alpha ,\eta_\beta ,\zeta_\gamma )\rightarrow (-\zeta_\gamma ,\xi_\alpha ,\eta_\gamma ) $. 
If we repeat the same analysis, 
then we can obtain the elliptic genus with the replacements $(N_2 ,N_3 ,N_1 )\rightarrow (N_3 ,N_1 ,N_2 )$ and
$(-\zeta_\gamma ,\xi_\alpha ,\eta_\gamma ) \rightarrow (-\eta_\beta ,-\zeta_\gamma , \xi_\alpha )  $. 
Thus we have confirmed the $(0,2)$ triality for the elliptic genus.

\subsubsection{Engineering the 2d $\mathcal{N}=(0,2)$ SQCD from 4d SQCD }
\begin{table}[tbp]
\begin{center}
  \begin{tabular}{|c|c  c c c |}
  \hline                                & $U(N_c)$       &$U(1)_R$   & $SU(N_1 )$ &  $SU(2)$ \\
\hline $Q_{i=1,\cdots ,N_1}$  & fund             & 0             & a-fund        &   1      \\  
         $\tilde{Q}_1$               & a-fund         &  $1-N_2$   & 1             &  1 \\  
         $\tilde{Q}_2$                  & a-fund         &  $1+N_3$   & 1             &  1 \\ 
         $\tilde{Q}_{i=3,\cdots ,N_1}$  & a-fund     &  $1$     & 1             &  1 \\   
        $M_{i=1,\cdots ,N_1} $                     & 1              &     $1+N_2$          &  fund          &1 \\  
      $\Omega_{s=1,2}$                    & det                & 2          &    1                                &fund  \\ \hline
         fugacities                    &-               & -             &   $\xi_\alpha$ &   $\lambda_s$        \\ 
        magnetic flux               &-               & 1              & 0                  &   0 \\ \hline
  \end{tabular}
\end{center}
\caption{Matter content of the 4d SQCD giving the elliptic genus of the 2d $\mathcal{N}=(0,2)$ SQCD. 
This theory has the rank $N_c =(N_1 +N_2 -N_3 )/2$ and superpotential $W=M_i \tilde{Q}_1 Q_i $.}
\label{tab:4dSQCD}
\end{table}
We discuss that 
there are 4d SQCDs on $T^2 \times S^2$ giving
the elliptic genus of the 2d $\mathcal{N}=(0,2)$ SQCD described in tab.~\ref{tab:2dSQCD}.
Let us consider the 4d $\mathcal{N}=1$ $U(N_c )$ SQCD with
\begin{itemize}
\item $N_1$ fundamental multiplets $Q_i$ with R-charge $r_f^{(i )}$

\item $N_1$ anti-fundamental multiplets $\tilde{Q}_i$ with R-charge $r_a^{(i)}$

\item 2 chiral multiplets $\Omega_s$ in det representation with R-charge 2

\item $N_1$ singlet chiral multiplets $M_i$ with R-charge $1+N_2$ and the superpotential $W=M_i \tilde{Q}_1 Q_i $

\end{itemize}
We have included the matters in the det representation in order to
cancel mixed anomaly between $U(1)_R$ and $U(1)$ part of the gauge group.
Conditions for all the gauge anomaly cancellations are boiled down to the following single equation
\begin{\eq}
 \sum_{i=1}^{N_1} ( r_f^{(i)}  +r_a^{(i)} )  =2N_1  -2N_c .
\end{\eq}
Because we do not turn on magnetic flux of flavor symmetries here,
we have to take the R-charges to be integers satisfying this condition.
Here let us take the R-charges as (the setup is summarized in tab.~\ref{tab:4dSQCD})
\begin{\eq}
r_f^{(i)}=0 ,\quad r_a^{(1)}=1-N_2 ,\quad r_a^{(2)}=1+N_3 , \quad r_a^{(i)}=1\ \  (i=3,\cdots ,N_1 ) .
\end{\eq}
Recalling our formula,
we easily see that
corresponding zero mode theory of the 4d SQCD on $T^2 \times S^2$
is the 2d $\mathcal{N}=(0,2)$ SQCD on $T^2$.
Indeed there are simple correspondences among the one-loop determinants:
\begin{\eqa}
&&\prod_{i=1}^{N_1}Z_{Q_i} =Z_P ,\quad 
Z_{\tilde{Q}_1} = \left. Z_\Phi \right|_{\eta_\beta =\{ m\sigma \}}  ,\quad 
Z_{\tilde{Q}_2} = \left. Z_\Psi \right|_{\zeta_\gamma =\{ m\sigma \}} ,\NN\\
&& Z_{\tilde{Q}_{i=3,\cdots N_1}} =1 ,\quad
\prod_{i=1}^{N_1}Z_{M_i} = \left. Z_\Gamma \right|_{\eta_\beta =\{ m\sigma \}} .
\end{\eqa}
Thus, the $T^2 \times S^2$ partition function of the 4d SQCD 
is the same as the elliptic genus of the 2d $\mathcal{N}=(0,2)$ SQCD with the special fugacities.

\begin{table}[tbp]
\begin{center}
  \begin{tabular}{|c|c  c c c |}
  \hline                                & $U(N_1 -N_c)$       &$U(1)_R$   & $SU(N_1 )$ &  $SU(2)$ \\
\hline $Q_{i=1,\cdots ,N_1}'$  & a-fund             & 0             & fund        &   1      \\  
         $\tilde{Q}_1'$               & fund         &  $1+N_2$   & 1             &  1 \\  
         $\tilde{Q}_2'$                  & fund         &  $1-N_3$   & 1             &  1 \\ 
         $\tilde{Q}_{i=3,\cdots ,N_1}'$  & fund     &  $1$     & 1             &  1 \\   
        $M_{i=1,\cdots ,N_1}' $                     & 1              &     $1+N_3$          &  a-fund          &1 \\  
      $\Omega_{s=1,2}'$                    & det                & 2          &    1                                &fund  \\ \hline
         fugacities                    &-               & -             &   $\xi_\alpha$ &   $\lambda_s$        \\ 
        magnetic flux               &-               & 1              & 0                  &   0 \\ \hline
  \end{tabular}
\end{center}
\caption{Matter content for the Seiberg dual of the 4d SQCD in tab.~\ref{tab:4dSQCD},
which gives the elliptic genus for the triality pair of the 2d $\mathcal{N}=(0,2)$ SQCD. 
This theory has the rank $(N_3 +N_1 -N_2 )/2$ and superpotential $W=M_i' \tilde{Q}_1' Q_i' $.}
\label{tab:dual4dSQCD}
\end{table}
As in usual Seiberg duality \cite{Seiberg:1994pq},
let us consider the 4d $U(N_1 -N_c )$ SQCD with the matter content listed in tab.~\ref{tab:dual4dSQCD}.
Then we easily find that
the partition function of this theory is the same as
the elliptic genus of the 2d SQCD with $(N_1 ,N_2 ,N_3 ) \rightarrow  (N_3 ,N_1 ,N_2 )$ and
$(\xi_\alpha ,\eta_\beta ,\zeta_\gamma ) \rightarrow (\{ m\sigma \} ,\xi_\alpha ,\{ m\sigma \}) $.
Since the 2d SQCD elliptic genus enjoys the $(0,2)$ triality,
this indicates that 
the 4d SQCD partition function on $T^2 \times S^2$ also enjoys the Seiberg duality.

\begin{table}[t]
\begin{center}
  \begin{tabular}{|c|c  c c c |}
  \hline                                & $U(N_c)$       &$U(1)_R$   & $SU(N_2 )$ &  $SU(2)$ \\
\hline                     $Q_{1}$  & fund             & $1-N_1$           & 1        &   1      \\  
                          $Q_{2}$  & fund               &  $1+N_3$           & 1        &   1      \\  
$Q_{i=3,\cdots ,N_2}$        & fund                & 1                      & 1       &   1      \\  
    $\tilde{Q}_{i=1,\cdots ,N_2}$  & a-fund     &  $0$                & fund             &  1 \\   
        $M_{i=1,\cdots ,N_1} $     & 1              &     $1+N_1$       &  a-fund          &1 \\  
      $\Omega_{s=1,2}$                    & det                & 2          &    1                                &fund  \\ \hline
         fugacities                    &-               & -             &   $\xi_\alpha$ &   $\lambda_s$        \\ 
        magnetic flux               &-               & 1              & 0                  &   0 \\ \hline
  \end{tabular}
\end{center}
\caption{Matter content of another 4d SQCD giving the elliptic genus of the 2d $\mathcal{N}=(0,2)$ SQCD. 
This is $U(N_c )$ gauge theory with $N_2$ flavors.}
\label{tab:4dSQCD2}
\end{table}
How can we get the remaining part of the triality $(N_1 ,N_2 ,N_3 ) \rightarrow  (N_2 ,N_3 ,N_1 )$?
For this purpose, let us consider the 4d $U(N_c )$ SQCD with $N_2$ flavors described in tab.~\ref{tab:4dSQCD2}.
We easily find that
this theory gives the same partition function as the 4d theory in tab.~\ref{tab:4dSQCD} and
elliptic genus of the 2d SQCD in tab.~\ref{tab:2dSQCD}.
If we consider the Seiberg dual of this theory in tab.~\ref{tab:dual4dSQCD2},
then its zero-mode theory becomes the 2d SQCD with $(N_1 ,N_2 ,N_3 ) \rightarrow  (N_2 ,N_3 ,N_1 )$ and
hence the 4d partition function is the same as the elliptic genus of the 2d SQCD.
Thus the 2d $(0,2)$ triality guarantees the 4d duality for the partition function on $T^2 \times S^2$ and
the $(0,2)$ triality for the elliptic genus comes from the Seiberg duality for the partition function on $T^2 \times S^2$.
It is interesting 
if we further test this in other observables or 
find more physical arguments as in connection between 4d and 3d dualities \cite{Aharony:2013dha}.

\begin{table}[t]
\begin{center}
  \begin{tabular}{|c|c  c c c |}
  \hline                                & $U(N_2 -N_c)$       &$U(1)_R$   & $SU(N_2 )$ &  $SU(2)$ \\
\hline                     $Q_{1}'$  & fund             & $1+N_1$           & 1        &   1      \\  
                          $Q_{2}'$  & fund               &  $1-N_3$           & 1        &   1      \\  
$Q_{i=3,\cdots ,N_2}'$        & fund                & 1                      & 1       &   1      \\  
  $\tilde{Q}_{i=1,\cdots ,N_2}'$  & a-fund     &  $0$                & a-fund             &  1 \\   
        $M_{i=1,\cdots ,N_2}' $     & 1              &     $1+N_1$       &  fund          &1 \\  
      $\Omega_{s=1,2}'$                    & det                & 2          &    1                                &fund  \\ \hline
         fugacities                    &-               & -             &   $\xi_\alpha$ &   $\lambda_s$        \\ 
        magnetic flux               &-               & 1              & 0                  &   0 \\ \hline
  \end{tabular}
\end{center}
\caption{The Seiberg dual of the 4d SQCD described in tab.~\ref{tab:4dSQCD2}
giving the elliptic genus of the 2d $\mathcal{N}=(0,2)$ SQCD with $(N_1 ,N_2 ,N_3 )\rightarrow (N_2 ,N_3 ,N_1 )$.
This theory has the rank $(N_2 +N_3 -N_1 )/2$ and the superpotential $W=M_i' \tilde{Q}_i' Q_1' $.
}
\label{tab:dual4dSQCD2}
\end{table}

\section{Conclusion and discussions}
\label{sec:conclusion}
In this paper we have studied the partition function of 4d $\mathcal{N}=1$ supersymmetric gauge theory on $T^2 \times S^2$.
We have shown by supersymmetry localization that
the partition function on $T^2 \times S^2$ is given by elliptic genus of 2d $\mathcal{N}=(0,2)$ gauge theory 
and obtained the exact formula.
This result is natural extension of the previous study \cite{Closset:2013sxa} in theory with only chiral multiplets.
Although \cite{Nishioka:2014zpa} also discussed theories with vector multiplets,
they did not taken the gaugino zero mode into account and did not obtained final formula.
We have appropriately treated the gaugino zero mode
by relating our analysis to the careful analysis of the elliptic genus \cite{Benini:2013xpa}.

Our result shows that
if we consider certain 4d SUSY gauge theory on $T^2 \times S^2$,
then we have corresponding 2d SUSY gauge theory on $T^2$, 
which gives the same partition function.
This fact enables us to find 
nontrivial relations between properties of 4d and 2d supersymmetric gauge theories.
Indeed we have shown that
the 2d $(0,2)$ triality \cite{Gadde:2013lxa} for the elliptic genus
comes from the 4d Seiberg duality \cite{Seiberg:1994pq} 
for the partition function on $T^2 \times S^2$.
Another possible attractive direction, which we have not pursued here, is symmetry.
We have shown that
the $T^2 \times S^2$ partition function is given only by the zero-modes along $S^2$,
which is described by the 2d $\mathcal{N}=(0,2)$ supersymmetric theory.
This fact implies that
the sub-sector of the 4d $\mathcal{N}=1$ theory would have hidden infinite dimensional symmetry at infrared fixed point.
It is interesting 
if we can relate our result to recent arguments on hidden symmetries in 4d \cite{Beem:2013sza,Strominger:2013lka,He:2014cra,He:2015zea}.
Also, some 2d CFTs have higher spin symmetry and are expected to be dual to Vasiliev theory on $AdS_3$ (see e.g. \cite{Gaberdiel:2012uj}).
It would be illuminating
if we can engineer 4d theories on $T^2 \times S^2$,
which give elliptic genera of 2d supersymmetric CFTs with higher spin symmetries \cite{Creutzig:2013tja,Gaberdiel:2013vva,Creutzig:2014ula,Gaberdiel:2014cha}.

It would be interesting to study the partition function of $\mathcal{N}=2$ Gaiotto theory \cite{Gaiotto:2009we} on $T^2 \times S^2$,
which is obtained by compactification of the 6d $\mathcal{N}=(2,0)$ theory on Riemann surface, and
find $T^2 \times S^2$ version of the AGT relation \cite{Alday:2009aq}.
For this purpose, previous studies on 4d superconformal index \cite{Kinney:2005ej} would be helpful.
It is known that
the 4d superconformal indices of class $S$ theories 
correspond to correlation functions of 2d TQFT on the Riemann surfece \cite{Rastelli:2014jja}.
We expect that the $T^2 \times S^2$ partition functions of the class $S$ theories
have also similar structures.

One of remaining questions is on factorization of supersymmetric partition functions \cite{Pasquetti:2011fj}.
It is known that
partition functions of supersymmetric theories on some spaces exhibit
structures like Heegaard decomposition of the spaces.
For example,
3d $\mathcal{N}=2$ theories on squashed $S^3$, $S^2 \times S^1$ and $S^3 /\mathbb{Z}_k$
can be interpreted as particular gluings of partition functions 
on $D^2 \times S^1$ \cite{Beem:2012mb,Nieri:2013yra,Yoshida:2014ssa}.
There is much evidence for this obtained by integration of Coulomb branch localization formula \cite{Pasquetti:2011fj,Taki:2013opa,Hwang:2012jh,Imamura:2013qxa}
and direct derivation by Higgs branch localization \cite{Fujitsuka:2013fga,Benini:2013yva} for some theories 
(see also another argument \cite{Alday:2013lba} and similar structures in 2d \cite{Benini:2012ui,Doroud:2012xw,Hori:2013ika,Honda:2013uca}).
Similar structure also appears on 4d superconformal index,
which is partition function on $S^3 \times S^1$ \cite{Yoshida:2014qwa,Peelaers:2014ima}.
It is natural to wonder
if the partition function on $T^2 \times S^2$ 
can be also interpreted as a gluing of partition functions on $T^2 \times D^2$.
Indeed if we further add the deformation term
\begin{\eq}
\mathcal{L}_H = \delta \Bigl[ {\rm Tr} (\zeta^\dag\lam -\tzeta^\dag\tlam ) h(\phi ) \Bigr] ,\quad 
{\rm with}\  h(\phi ) =\frac{i}{2}\left( \phi \tilde{\phi}-{\rm const.}\times \mathbf{1}\right) ,
\end{\eq}
then bosonic part of $\mathcal{L}_{\rm vec}$ plus $\mathcal{L}_H$ is given by
\begin{\eq}
\left. \mathcal{L}_{\rm vec} + \mathcal{L}_H \right|_{\rm bos.}
=\frac{1}{2}\mathcal{F}_{ip}^2 +\frac{1}{4}\mathcal{F}_{pq}^2
 +\frac{1}{2}(\mathcal{F}_{12} +ih(\phi ))^2 -\frac{1}{2}(D-ih(\phi ))^2 .
 \label{HiggsQ}
\end{\eq}
Combined with the action of the chiral multiplet,
we can show that the saddle point of localization is described by
supersymmetric vortex solution with {\it one} supercharge. 
Therefore we can perform Higgs branch localization by using this deformation term in principle. 
However one of the vortex equations given by (\ref{HiggsQ}) exists not only on the north or south pole on $S^2$, but 
also exists on every point on $S^2$.  Moreover the vortex preserves only one supercharge.  
These are different from usual story of the factorization,
where the vortex equation (anti-vortex equation)  preserves {\it two} supercharges,
appears only on the north pole (the south pole) in saddle point
and each of their world volume theory is captured by 2d $\mathcal{N}=(0,2)$ theory or its dimensional reductions. 
Since the vortex preserves only one supercharge for our $T^2 \times S^2$ case,
its world volume theory seems to be 2d $\mathcal{N}=(0,1)$ theory,
whose partition function has been less studied.
Thus, if factorization occurs also for $T^2 \times S^2$,
then its structure would be slightly different from the usual story.
It is interesting to further pursue this direction.

\subsection*{Acknowledgment}
We are grateful to Masashi Fujitsuka for his early collaboration and many discussions.
We would also like to thank Hee-Joong Chung,  
Abhijit Gadde, Dileep Jatkar, Kimyeong Lee, Sara Pasquetti, Yuji Tachikawa and Piljin Yi
for valuable discussions.
M.H. also thank the members of the HRI string group 
to attend his long seminar, 
where a part of this work was presented.

\appendix
\section{Convention}
Here we summarize our convention of spinors, which is based on \cite{Dumitrescu:2012ha,Assel:2014paa}. 
We have two Weyl spinors in representations of the rotational group $SO(4)=SU(2)_+ \times SU(2)_-$: 
$\zeta_\alpha$ in $SU(2)_+$ doublet and $\tzeta^{\dalpha}$ ($\dalpha=1,2$) in $SU(2)_-$ doublet with $\alpha ,\dalpha =1,2$.
We define contraction, upper and lower indices as
\begin{\eq}
\zeta\chi =\zeta^\alpha \chi_\alpha ,\quad
\tzeta\tilde{\chi} =\tzeta_{\dalpha} \tilde{\chi}^{\dalpha} ,  \quad
\zeta_\alpha = \eps^{\alpha\beta}\zeta_\beta ,\quad
\tzeta_{\dalpha} = \eps_{\dalpha\dbeta} \tzeta^{\dalpha} ,
\end{\eq}
with $\eps^{12}=+1 ,\quad \eps_{12}=-1$.
We also define Hermitian conjugates as
\begin{\eq}
(\zeta^\dag )^\alpha = ( \zeta_\alpha )^\ast , \quad 
(\tzeta^\dag )_{\dalpha} = ( \tzeta^{\dalpha} )^\ast .
\end{\eq}
The sigma matrices are given by
\begin{\eq}
\sigma^{\hat{\mu}}_{\alpha\dalpha} = (\vec{\sigma},-i), \quad \tsigma^{\hat{\mu}\dalpha\alpha} = (-\vec{\sigma},-i),
\end{\eq}
where $\hat{\mu},\hat{\nu}=1,\cdots,4$ are local Lorentz indices and $\vec{\sigma}$ denotes the Pauli matrices. 
These satisfy the following useful identities
\begin{\eqa}
&&\sigma_{\hat{\mu}} \tsigma_{\hat{\nu}} +\sigma_{\hat{\nu}} \tsigma_{\hat{\mu}} =-2\delta_{\hat{\mu}\hat{\nu}},\quad
\tsigma_{\hat{\mu}} \sigma_{\hat{\nu}} +\tsigma_{\hat{\nu}} \sigma_{\hat{\mu}} =-2\delta_{\hat{\mu}\hat{\nu}} ,\NN\\
&&\frac{1}{2}\eps_{\hat{\mu}\hat{\nu}\hat{\rho}\hat{\lambda}} \sigma^{\hat{\rho}\hat{\lambda}} =\sigma_{\hat{\mu}\hat{\nu}} ,\quad
\frac{1}{2}\eps_{\hat{\mu}\hat{\nu}\hat{\rho}\hat{\lambda}} \tsigma^{\hat{\rho}\hat{\lambda}} =-\tsigma_{\hat{\mu}\hat{\nu}}, 
\end{\eqa}
where $\eps_{1234}=1$ and 
\begin{\eq}
\sigma_{\hat{\mu}\hat{\nu}} = \frac{1}{4}(\sigma_{\hat{\mu}} \tsigma_{\hat{\nu}}  -\sigma_{\hat{\nu}} \tsigma_{\hat{\mu}} ) ,\quad
\tsigma_{\hat{\mu}\hat{\nu}} = \frac{1}{4}(\tsigma_{\hat{\mu}} \sigma_{\hat{\nu}}  -\tsigma_{\hat{\nu}} \sigma_{\hat{\mu}} ) .
\end{\eq}

\section{Eta function and theta function}
\label{sec:theta}
Here we briefly summarize properties of the Dedekind eta function and Jacobi theta function.
\subsection{Eta function}
The Dedekind eta function is defined by
\begin{\eq}
\eta (\tau ) = q^{\frac{1}{24}} \prod_{n=1}^\infty (1-q^n ) ,
\end{\eq}
where
\begin{\eq}
q =e^{2\pi i\tau} .
\end{\eq}
This has the following properties
\begin{\eq}
\eta (\tau +1)=e^{\frac{i\pi}{12}}\eta( \tau ) ,\quad
\eta (-1/\tau ) = \sqrt{-i\tau}\eta (\tau ).
\end{\eq}

\subsection{Theta function}
The Jacobi theta function is defined by
\begin{\eq}
\theta_1 (\tau |z)
=-iq^{\frac{1}{8}}y^{\frac{1}{2}} \prod_{k=1}^\infty (1-q^k )(1-yq^k ) (1-y^{-1}q^{k-1}) ,
\end{\eq}
where
\begin{\eq}
y=e^{2\pi iz} .
\end{\eq}
This satisfies some transformation properties
\begin{\eqa}
&&\theta_1 (\tau |z+a+b\tau )
=(-1)^{a+b}e^{-2\pi b z -i\pi b^2 \tau}\theta_1 (\tau | z) \quad
{\rm with}\ a,b\in\mathbb{Z}  ,\NN\\
&&\theta_1 (\tau +1 |z )  =e^{\frac{\pi i}{4}} \theta_1 (\tau |z) ,\quad
\theta_1 \left( -\frac{1}{\tau}  | \frac{z}{\tau} \right)  =-i\sqrt{-i\tau} e^{\frac{\pi iz^2 }{\tau}} \theta_1 (\tau |z) .
\end{\eqa}
The following formula is useful for picking up residues
\begin{\eq}
\frac{1}{2\pi i}\oint_{u=a+b\tau} du \frac{1}{\theta_1 (\tau |u )}
= \frac{(-1)^{a+b}e^{i\pi b^2 \tau}}{2\pi\eta^3 (\tau )}\quad
{\rm with}\ a,b\in\mathbb{Z} .
\end{\eq}

\section{Zeta function regularization and scheme dependent factor}
\label{app:reg}
The one-loop determinants are expressed in terms of   the following infinite product 
\begin{eqnarray}
\prod_{n_1, n_2 \in \mathbb{Z}} (n_1+n_2 \tau+z). 
\label{unregtheta}
\end{eqnarray}
Since this infinite product is divergent, we have to specify a regularization scheme. We evaluate the infinite product by 
the two different ways and see the scheme dependent factor.

First, we split the infinite product as 
\begin{eqnarray}
\prod_{n_1, n_2 \in \mathbb{Z}} (n_1+n_2 \tau+z)
&=&\prod_{n_1 \in \mathbb{Z} } (n_1 \tau +z) \prod_{n_2=1}^{\infty} (-n^2_2) \left[1-\frac{(n_1 \tau+ z)^2}{n^2_2} \right] \nonumber \\
&=&\prod_{n_1 \in \mathbb{Z} }  2(-1)^{\frac{1}{2}} \sin \pi (n_1 \tau+ z) 
\end{eqnarray}
From the first line to the second line in the above equation, 
we have used $\pi z\prod_{n \in \mathbb{Z}} (1-z^2/n^2)=\sin \pi z$ 
and $e^{\sum_{n=1}^{\infty} \log (-n^2)}=e^{\zeta(0) \log(-1)-2\zeta^{'}(0)}=2\pi (-1)^{-\frac{1}{2}}$.
We split the above divergent infinite product  as
\begin{eqnarray}
\prod_{n \in \mathbb{Z} }   {2(-1)^{\frac{1}{2}}} \sin \pi (n \tau+ z)= 
{2(-1)^{\frac{1}{2}}} \sin \pi z \prod_{n=1}^{\infty}  [ {(-1)^{\frac{1}{2}}} ]^2 e^{-2 \pi i n \tau} (1-e^{2\pi i n \tau} e^{2\pi iz}) 
(1-e^{2\pi i n \tau} e^{-2\pi iz}) \nonumber \\
 \end{eqnarray}
Again we have used the zeta function regularization $\prod_{n=1}^{\infty} x^2=e^{\zeta (0) \log x^2} =x^{-1}$ and $\prod_{n=1}^{\infty} e^{ nx} =e^{x \zeta(-1)}=e^{\frac{x}{12}}$. Then we obtain a regularized infinite product as
 \begin{eqnarray}
\prod_{n,m \in \mathbb{Z}} (n+m\tau+z)=\frac{\theta_1(\tau |z)}{\eta(\tau)}.
\end{eqnarray}
In our paper, we adapt this regularization scheme  and  also include fugacity independent 
over all constant $i$ in front of  the one-loop determinant  which is not fixed by the localization argument.  
 
Next, we evaluate the infinite product by another regularization scheme. 
We introduce the following regularized infinite products by the double gamma function $\Gamma_2(z|\omega_1,\omega_2)$ or
 double zeta function $\zeta(s,z|\omega_1,\omega_2)$ as
\begin{eqnarray}
\prod_{n_1, n_2 \ge 0} (n_1\omega_1+n_2 \omega_2+z):=
\Gamma_2(z|\omega_1,\omega_2)^{-1}
:=\exp (- \zeta^{'} (s=0,z|\omega_1,\omega_2))
\end{eqnarray}
with
\begin{eqnarray}
\zeta(s,z|\omega_1,\omega_2):=\sum_{n_1,n_2 \ge 0} (n_1\omega_1+n_2 \omega_2+z)^{-s}.
\end{eqnarray}
Then we define a regularization of (\ref{unregtheta}) in terms of the double gamma function as
\begin{eqnarray}
\prod_{n_1, n_2 \in \mathbb{Z}} (n_1+n_2 \tau+z) &=&\left[ \Gamma_2(z|1,\tau) \Gamma_2(z-\tau|1,-\tau)  
\Gamma_2(1-z|1,-\tau)  \Gamma_2(1+\tau-z|1,\tau) \right]^{-1} \nonumber \\
&=& e^{\frac{\pi i}{\tau} (z^2-z+\frac{1}{6}) } \frac{\theta_1 (\tau|z)}{\eta(\tau)}.
\label{regoneloop}
\end{eqnarray}
From the first line to the second line, we have used proposition 2 in \cite{shintani}.

In the second regularization scheme an additional  factor $e^{\frac{\pi i}{\tau}(z^2-z+\frac{1}{6})}$ appeared in  (\ref{regoneloop}).
This factor breaks invariance under the large gauge transformation or  integer shifts $u \to u +n, (n \in \mathbb{Z})$.  
Although, it is not clear that this factor have physical meaning or it can be eliminated by local counter terms in 4d,  
we briefly study the cancellation conditions of this anomalous factor.
The coefficients of quadratic terms  for fugacities $u$ and $\xi$ and $\sigma$   in the exponential   are proportional to 
\begin{eqnarray}
&&u^a u^b: \mathrm{Tr}_{\text{adj}}(H^a H^b) +\sum_{i} (\mathbf{r}_i-1) \mathrm{Tr}_{\mathbf{R}_i}(H^a H^b), 
\label{gganomaly} \\
&&u^a \xi^b:\sum_{i} (\mathbf{r}_i-1) \mathrm{Tr}_{\mathbf{R}_i}(H^a) q^{(i,b)}_f ,
\label{gfanomaly} \\
&& \xi^a \xi^b:  \sum_{i} (\mathbf{r}_i-1) q^{(i,a)}_f q^{(i,b)}_f, 
\label{ffanomaly} \\
&&\sigma^2:\sum_{i: \mathbf{r}_i>1}\sum^{\frac{\mathbf{r}_i}{2}-1}_{m=-\frac{\mathbf{r}_i}{2}+1} m^2 
-\sum_{i: \mathbf{r}_i<1} \sum^{\frac{|\mathbf{r}_i|}{2}}_{m=-\frac{|\mathbf{r}_i|}{2}} m^2. 
\label{ssanomaly}
\end{eqnarray}
Here $\{ H^a \}_{a=1, \cdots, |G|}$ is the generator of Cartan subalgebra of Lie algebra of $G$. 
(\ref{gganomaly}) and (\ref{gfanomaly}) are same as  the coefficients of gauge-gauge  and gauge-flavor anomaly, respectively   
and also (\ref{ffanomaly}),  (\ref{ssanomaly}) are the one of flavor-flavor anomalies.  
When the gauge-gauge and gauge-flavor anomalies in two dimensions are canceled, the one-loop determinant is invariant under the integer shift. 
Next linear terms in the exponential is written as  
\begin{eqnarray}
&&u^a: \sum_{i} (\mathbf{r}_i-1) \mathrm{Tr}_{\mathbf{R}_i}(H^a ), \quad \xi^a :  \sum_{i} (\mathbf{r}_i-1)  q^{(i,a)}_f .
\label{danomaly}
\end{eqnarray}
The coefficient of $u^a$ in (\ref{danomaly}) is proportional to axial anomaly in two dimensions. 
The fugacity independent  factor is  given by $e^{\frac{\pi i}{6\tau}(|G|^2+\sum_{i} (\mathbf{r}_i-1)|\mathbf{R}_i|)}$.

\section{Monopole spherical harmonics on $S^2$}
\label{sec:harmonics}
In this appendix,
we briefly summarize properties of monopole spherical harmonics on $S^2$.
\subsection{Scalar Monopole spherical harmonics}
Laplacian on $S^2$ is given by
\begin{\eq}
\Delta_{S^2} = -(1+z\bz )^2 \del_z \del_{\bz} -\frac{r}{2}(1+z\bz )\left( z\del_z -\bz\del_{\bz} -\frac{r}{2} \right) -\frac{r^2}{4}.
\end{\eq}
The scalar monopole spherical harmonics $ Y_{rJm}$ satisfies
\begin{\eqa}
&&\Delta_{S^2} Y_{rJm} = \left( j(j+1) -\frac{r^2}{4} \right) Y_{rJm} ,\quad
J^2 Y_{rJm} =j(j+1)Y_{rJm} ,\quad J_3 Y_{rJm} = m Y_{rJm} ,\NN\\
&&\int dz d\bz \sqrt{g_{S^2}} Y_{rJm}^\dag Y_{r J^\prime m^\prime} 
=\delta_{JJ^\prime} \delta_{mm^\prime},
\end{\eqa}
where 
\begin{\eq}
J=\frac{|r|}{2}, \frac{|r|}{2}+1 ,\cdots .
\end{\eq}

\subsection{Spinor Monopole spherical harmonics}
Dirac operator on $S^2$ in our notation is
\begin{\eqa}
\left( \tsigma^1 D_1 +\tsigma^2 D_2 \right)^{\dalpha\alpha}
= -\begin{pmatrix}  0 &  (1+z\bz )\del_{\bz} +\frac{r-2}{2}z  \cr
                          (1+z\bz )\del_z -\frac{r}{2}\bz  & 0  \end{pmatrix} .
\end{\eqa}
Then, the spinor harmonics satisfy 
\begin{\eq}
\left( \tsigma^1 D_1 +\tsigma^2 D_2 \right)^{\dalpha\alpha}
\psi_{r-1Jm}^\pm 
=\pm i \sqrt{\left( J+\frac{1}{2}\right)^2 -\frac{(r-1)^2}{4}}   \psi_{r-1Jm}^\pm ,
\end{\eq}
where $J=\frac{|r|}{2}-\frac{1}{2}, \frac{|r|}{2}+\frac{1}{2} ,\cdots ,$ and
\begin{\eq}
( \psi_{r-1Jm}^\pm )_\alpha 
= \frac{1}{\sqrt{2}} \begin{pmatrix}  \pm Y_{rJm}  \cr -iY_{r-2 Jm} \end{pmatrix} .
\end{\eq}
This relation leads the useful identities
\begin{\eqa}
&&\left( \tsigma^1 D_1 +\tsigma^2 D_2 \right)
\begin{pmatrix} 0\cr Y_{r-2 Jm}  \end{pmatrix} 
= -\sqrt{\left( J+\frac{1}{2}\right)^2 -\frac{(r-1)^2}{4}}  \begin{pmatrix}  Y_{rJm} \cr 0 \end{pmatrix}  ,\NN\\
&&\left( \tsigma^1 D_1 +\tsigma^2 D_2 \right)
      \begin{pmatrix} Y_{rJm} \cr 0 \end{pmatrix} 
= \sqrt{\left( J+\frac{1}{2}\right)^2 -\frac{(r-1)^2}{4}}  \begin{pmatrix} 0 \cr Y_{r-2 Jm}  \end{pmatrix} .
\end{\eqa}

\subsection{Vector spherical harmonics}
Here we need only usual vector spherical harmonics.
According to notation of \cite{Honda:2013uca},
the harmonics satisfy
\begin{\eqa}
&& \Delta_{S^2} C_{i,Jm}^\rho =-\left( J(J+1) -1 \right) C_{i,Jm}^\rho   ,\quad
 D^{i(0)}C_{i,Jm}^1 = -\sqrt{J(J+1) } Y_{0Jm} ,\NN\\
&&D^{i(0)}C_{i,Jm}^2 =0 ,\quad  D_i^{(0)} \eps^{ij} C_{j,Jm}^1 = 0 ,\quad D_i^{(0)} \eps^{ij} C_{j,Jm}^2 = -\sqrt{J(J+1) } Y_{0Jm} ,\NN\\
&&\int dz d\bz \sqrt{g_{S^2}}  g_{S^2}^{ij} C_{i,Jm}^{\rho} C_{j,J^\prime m^\prime}^{\rho^\prime}
= \delta_{\rho \rho^\prime} \delta_{J J^\prime} \delta_{mm^\prime} .
\end{\eqa}

\section{2d supersymmetric gauge theory on $T^2$}
In this appendix we write down
actions and supersymmetric transformations of
2d $\mathcal{N}=(2,2)$ vector multiplet, $\mathcal{N}=(2,2)$ chiral multiplet,
$\mathcal{N}=(0,2)$ vector multiplet, $\mathcal{N}=(0,2)$ chiral multiplet and $\mathcal{N}=(0,2)$ Fermi multiplet on $T^2$.
\subsection{$\mathcal{N}=(2,2)$ vector multiplet}
Lagrangian for $\mathcal{N}=(2,2)$ SYM
can be obtained by dimensional reduction of 4d $\mathcal{N}=1$ SYM on flat space
along $(1,2)$-direction:
\begin{\eq}
\calL_{\rm vec}
= {\rm Tr}\Biggl[ \frac{1}{4} \calF^{\mu\nu}\calF_{\mu\nu} -\frac{1}{2}D^2
+\frac{i}{2}\lam \sigma^\mu D_\mu \tlam +\frac{i}{2}\tlam \tsigma^\mu D_\mu \lam \Biggr] ,
\end{\eq}
where $\calF_{ij} = -i [A_i ,A_j ] $, $F_{pi} = D_p A_i$, $D_i (\cdot ) = - i [A_i ,(\cdot )]$.
The action is invariant under  
\begin{\eqa}
\delta \calA_\mu &=& i\zeta\sigma_\mu\tlam +i\tzeta\tsigma_\mu \lam ,\NN\\
\delta\lam &=& \calF_{\mu\nu}\sigma^{\mu\nu}\zeta +iD\zeta ,\NN\\
\delta\tlam &=& \calF_{\mu\nu}\tsigma^{\mu\nu}\tzeta -iD\tzeta ,\NN\\
\delta D &=& -\zeta\sigma^\mu  D_\mu \tlam     +\tzeta \tsigma^\mu  D_\mu \lam  .
\end{\eqa}

\subsection{$\mathcal{N}=(2,2)$ chiral multiplet}
The Lagrangian for $\mathcal{N}=(2,2)$ chiral multiplet
is also obtained by dimensional reduction of 4d $\mathcal{N}=1$ chiral multiplet:
\begin{\eqa}
\mathcal{L}_{\rm chi}
= D_\mu \tphi D^\mu \phi   +\tphi D\phi -\tF F 
 +i\tpsi\tsigma^\mu D_\mu \psi     +i\sqrt{2}(\tphi\lambda\psi -\tpsi\tlam\phi ) ,
\end{\eqa}
and the supersymmetric transformation is
\begin{\eqa}
\delta\phi &=& \sqrt{2}\zeta\psi ,\NN\\
\delta\psi &=& \sqrt{2}F\zeta +i\sqrt{2}(\sigma^\mu \tzeta )D_\mu \phi ,\NN\\
\delta F &=& i\sqrt{2}\tzeta\tsigma^\mu D_\mu \psi   -2i(\tzeta\tlam )\phi ,\NN\\
\delta\tphi &=& \sqrt{2}\tzeta\tpsi ,\NN\\
\delta\tpsi &=& \sqrt{2}\tF\tzeta +i\sqrt{2}(\tsigma^\mu \zeta )D_\mu \tphi ,\NN\\
\delta \tF &=& i\sqrt{2}\zeta\sigma^\mu D_\mu \tpsi       +2i\tphi (\zeta\lambda ) .
\end{\eqa} 
We can also add superpotential term:
\begin{\eq}
\mathcal{L}_{\rm pt}
= F_i W_i +\tF \tilde{W}_i -\frac{1}{2}W_{ij}\psi^i \psi^j -\frac{1}{2}\tilde{W}_{ij}\tpsi^i \tpsi^j .
\end{\eq}

\subsection{$(0,2)$ SYM}
\label{app:02SYM}
We can get Lagrangian and SUSY transformation of $(0,2)$ SYM by taking
\begin{\eq}
\zeta_\alpha =\begin{pmatrix} 0\cr \zeta_+ \end{pmatrix} ,\quad
\tzeta^{\dalpha} =\begin{pmatrix} \tzeta_+ \cr 0 \end{pmatrix} ,\quad
\lambda_\alpha =\begin{pmatrix} 0\cr \lam_+ \end{pmatrix} ,\quad
\tlam^{\dalpha} =\begin{pmatrix} \tlam_+ \cr 0 \end{pmatrix} .
\end{\eq}
Then we find
\begin{\eq}
\calL_{\Upsilon}
= {\rm Tr}\Biggl[ \frac{1}{4} \calF^{pq}\calF_{pq} -\frac{1}{2}D^2 +\tlam_+ D_{\bw}\lam_+ \Biggr] ,
\end{\eq}
and
\begin{\eqa}
&&\delta \calA_w =   -2\zeta_+ \tlam_+ +2\tzeta_+ \lam_+ ,\quad
\delta \calA_{\bw} =  0,\NN\\
&&\delta\lam_+ = i\calF_{34}\zeta_+ +iD\zeta_+ ,\quad
\delta\tlam_+ = i\calF_{34}\tzeta_+ -iD\tzeta_+ ,\NN\\
&&\delta D = i\zeta_+   D_{\bw} \tlam_+     -i\tzeta_+ D_{\bw}\lam_+ .
\end{\eqa}

\subsection{$(0,2)$ decomposition of $\mathcal{N}=(2,2)$ chiral multiplet}
Let us take
\begin{\eq}
\zeta_\alpha =\begin{pmatrix} 0\cr \zeta_+ \end{pmatrix} ,\quad
\tzeta^{\dalpha} =\begin{pmatrix} \tzeta_+ \cr 0 \end{pmatrix} ,\quad
\lambda_\alpha =\begin{pmatrix} 0\cr \lam_+ \end{pmatrix} ,\quad
\tlam^{\dalpha} =\begin{pmatrix} \tlam_+ \cr 0 \end{pmatrix} ,
\end{\eq}
and decompose the matter fermions as
\begin{\eq}
\psi_\alpha =\begin{pmatrix} \psi_- \cr \psi_+ \end{pmatrix} ,\quad
\tpsi^{\dalpha} =\begin{pmatrix} \tpsi_+ \cr \tpsi_- \end{pmatrix} .
\end{\eq}
Then, we can decompose $\mathcal{N}=(2,2)$ chiral multiplet 
into $(0,2)$ chiral and Fermi multiplet.
Lagrangian for $(0,2)$ chiral multiplet is given by
\begin{\eqa}
\mathcal{L}_\Phi
&=& D_p \tphi D^p \phi  +\tphi D\phi 
 -\tpsi_- D_w \psi_-     +i\sqrt{2}(\tphi\lam_+\psi_- +\tpsi_- \tlam_+ \phi ) .
\end{\eqa}
This is invariant under
\begin{\eqa}
&&\delta\phi = +\sqrt{2}\zeta_+ \psi_- ,\quad 
\delta\tphi = +\sqrt{2}\tzeta_+ \tpsi_-  ,\NN\\
&&\delta\psi_- 
=\sqrt{2} \tzeta_+ D_{\bw} \phi ,\quad
\delta\tpsi_- 
= \sqrt{2} \zeta_+ D_{\bw} \tphi .
\end{\eqa} 
Lagrangian for Fermi multiplet without potential is
\begin{\eqa}
\mathcal{L}_{\rm fermi}
=+\tpsi_+ D_{\bw}\psi_+   -\tF F ,
\end{\eqa}
which is invariant under
\begin{\eqa}
&&\delta\psi_+ = \sqrt{2}F\zeta_+  ,\quad
\delta\tpsi_+ = \sqrt{2}\tF\tzeta_+  ,\NN\\
&&\delta F 
=+\sqrt{2}\tzeta_+ D_{\bw} \psi_+   ,\quad
\delta \tF 
=+\sqrt{2}\zeta_+ D_{\bw}\tpsi_+ .
\end{\eqa} 
We can also add potential in supersymmetric way:
\begin{\eqa}
\mathcal{L}_{\rm fermi}
=+\tpsi_+ D_{\bw}\psi_+  -\tF F +EF+\tilde{E}\tF -\psi_+ \psi_-^E -\tpsi_+ \tpsi_-^E ,
\end{\eqa}
where
\begin{\eq}
E=E(\phi ),\quad \tilde{E}=\tilde{E}(\tphi ),\quad
\psi_-^E = \frac{\del E}{\del \phi^i }\psi_-^i ,\quad
\tpsi_-^E = \frac{\del \tilde{E}}{\del \tphi^i }\tpsi_-^i .
\end{\eq}
Note that this potential terms themselves are $\delta$-exact:
\begin{\eq}
\delta (\psi_+ E ) =\sqrt{2}\zeta_+ (FE -\psi_+ \psi_-^E ),\quad
\delta (\tpsi_+ \tilde{E}) =\sqrt{2}\tzeta_+ (\tF\tilde{E} -\tpsi_+ \tpsi_-^E ) .
\end{\eq}
Hence elliptic genus should be independent of parameters in $E$ and $\tilde{E}$.
Redefining
\begin{\eq}
G=F-\tilde{E},\quad \tilde{G}=\tilde{F}-E ,
\end{\eq}
we rewrite the Lagrangian as
\begin{\eqa}
\mathcal{L}_{\rm fermi}
=+\tpsi_+ D_{\bw}\psi_+  -\tilde{G}G  +\tilde{E}E -\psi_+ \psi_-^E -\tpsi_+ \tpsi_-^E ,
\end{\eqa}
where supersymmetric transformation is given by
\begin{\eqa}
&&\delta\psi_+ = \sqrt{2}(G+\tilde{E})\zeta_+  ,\quad
\delta\tpsi_+ = \sqrt{2}(\tilde{G}+E)\tzeta_+  ,\NN\\
&&\delta G = +\sqrt{2}\tzeta_+ D_{\bw} \psi_+ -\sqrt{2}\tzeta_+ \tpsi_-^E   ,\quad
\delta \tilde{G} = +\sqrt{2}\zeta_+ D_{\bw}\tpsi_+ -\sqrt{2}\zeta_+ \psi_-^E .
\end{\eqa}

\providecommand{\href}[2]{#2}\begingroup\raggedright\endgroup

\end{document}